# A magnification-based multi-asperity (MBMA) model of rough contact where the Greenwood-Williamson and Persson theories meet

Xu Guo[1], Benben Ma and Yichao Zhu[1]

*State Key Laboratory of Structural Analysis for Industrial Equipment,*

*International Research Center for Computational Mechanics,*

*Department of Engineering Mechanics, Dalian University of Technology,*

*Dalian, 116023, P.R. China*

**Absract**

Contact analysis without adhesion is still a challenging problem, mainly owing to the multiscale and self-fractal characteristics of rough surfaces. Up to now, theories for analyzing contact behavior of rough surfaces in literature can be generally categorized into two groups: the asperity-based Hertz contact models initiated by Greenwood and Williamson (G-W model), which is shown more accurate under small indentation distance, and the magnification-based pressure diffusion theory initiated by Persson, which is shown to work well under full contact conditions. The aim of this paper is to propose a theoretical model that can effectively formulate the contact status of rough surfaces during the entire compression process. This is achieved by integrating the idea of magnification, or evolving resolution into an asperity representation of rough surfaces, and a magnification-based multi-asperity model is thus established. In the derived model, the originally complex contact problem is decomposed into a family of sub-problems each defined on a morphologically simpler contact islands. Benefiting from the explicit results given by Greenwood and Williamson, the proposed method is relatively easy for numerical implementation. Compared to other G-W type models, the proposed method has especially shown its strength in the computation of the contact area. Moreover, the G-W and Persson models are found well connected by the proposed method. For its validation, the proposed model is well compared with existing numerical, theoretical and experimental results. In particular, the proposed model has shown its excellency through comparison with representative theoretical, numerical and experimental data compiled in the "contact challenge" test by Mueser et al.

---

[1] Corresponding authors. guoxu@dlut.edu.cn (Xu Guo) and yichaozhu@dlut.edu.cn (Yichao Zhu)

## 1. Introduction

It is widely known that no surface is absolutely smooth. Even a macroscopically polished surface consists of many microscopically undulated sub-surfaces. Therefore, an accurate estimation of the actual contact area and the resulting contact force when a rough surface gets compressed, is of utmost importance for a wide range of engineering applications (e.g., structural adhesives, friction of tires, wear, lubrication and seals). Actually, one of the most challenging issues in the study of rough surface is how to effectively identify the morphology of a rough surface. In 1960s, Greenwood and Williamson (1966) described the morphology of a rough surface as a set of non-interacting spherical asperities. However, the calculations of the morphological parameters, as suggested by Whitehouse and Archard (1970), normally get highly affected by the resolution (a concept firstly proposed by Archard (1953)). Due to the limitation of the experimental resolution then available for G-W models, the number of detected asperities was far too small to contemplate any correlation between the mean curvature and height of a rough surface. Then by treating a rough surface as a random field, Nayak (1971) related all key parameters characterizing a rough surface to three profile moments (termed as the Nayak's moments), which are widely considered as the foundations of the theoretical study of surface roughness (Greenwood, 2007).

Nevertheless, the Nayak's moments, which can be further associated with the spectral density of a rough surface, depended heavily on the range of detected wave numbers. Recently, based on the fact that most natural and engineering surfaces are self-affine fractals (Ciavarella et al., 2004; Persson, 2014), power-law type spectral density functions are employed to calculate the mean summit curvature and asperity density (Bottiglione et al., 2009; Lorenz et al., 2010; Yastrebov et al., 2017). Under this set-up, although the mean summit curvature and asperity density are calculated by averaging over the whole spectrum, the resulting quantities are actually weighted to the high wave number end. To deal with the multi-scale feature of rough surfaces, Ciavarella et al. (2004) employed a Weierstrass series to discretely represent a two-dimensional fractal surface profile, which has achieved some improvements to the original G-W model (Ciavarella et al., 2004, 2006a). Later on, this treatment has also been successfully generalized into three-dimensional contact analysis (Ciavarella et al., 2006b; Papangelo el al., 2017). However, as far as the authors know, consensus on how to effectively represent a rough surface is still far from being reached.

Practically, most of the up-to-date methods for formulating the contact status at rough surfaces can be categorized into two groups: i) the multi-asperity contact approaches, in which the G-W model, initialized by Greenwood and Williamson (1966) is the most widely used one, and ii) the magnification-based pressure diffusion method introduced by Persson (2001).

In the multi-asperity contact approaches (Bush et al., 1975; Greenwood, 2006; Greenwood and Williamson, 1966; Nayak, 1971), the contact problem between rough surfaces is treated as a rigid and flat surface compressing an elastic solid equipped with a rough surface, whose complex topography is effectively represented by a set of non-interacting asperities of characterizing shape. Such treatment enables ones to analyze the contact behavior of each individual asperity by applying the classical Hertz theory (Hertz, 1881) to account for the compression of elastic asperities (Lorenz et al., 2010). Then the overall contact status, such as the actual contact area and the resulting contact force, can be calculated through summarizing the results from individual asperities. For example, in the classical G-W model, it is assumed that the constituting asperities of a rough surface are of identical curvature radius, and they follow a horizontally uniform and vertically Gaussian distribution. Enjoying its simplicity and tractability, the G-W approach lays a solid foundation for resolving a huge number of rough contact problems, e.g., adhesive contact analysis (Zhang et al., 2014), elasto-plastic contact analysis (Song et al., 2017; Song et al., 2016), electrical conduction (Ciavarella et al., 2008; Sevostianov and Kachanov, 2008), partial slip behavior (Jin et al., 2015), chemical mechanical polishing (Zhao and Chang, 2002) and unloading of contacting rough surfaces (Kadin et al., 2006; Kim et al., 2004). Given the great success of the G-W approach in engineering applications, it is, however, mainly a qualitative understanding and generally limited by the following two issues (which also apply to other multi-asperity contact models (Bush et al., 1975; Greenwood, 2006; Nayak, 1971)). Firstly, since it is established based on the Hertz theory, the assumption of small strain needs to be imposed. Secondly, the morphological parameters (e.g., the density and the radius of the asperities) used in the G-W model neglects the multiscale feature of the surface roughness (Carbone and Bottiglione, 2011; Thomas, 1998), and they are highly related to the number of sampling points according to which the surface topography is measured. Therefore, in general, G-W and the other related multi-asperity contact approaches are believed to be more applicable to small indentation distances or contact areas (Carbone and Bottiglione, 2008, 2011; Lorenz et al., 2010; Persson, 2006; Thomas, 1998).

In contrast to the G-W approach where rough surfaces are characterized by identical spheres, the magnification-based pressure diffusion method initiated by Persson (2001) (termed as the Persson model in this article) is proposed to take into account the multi-scale feature of rough surfaces. In his work, the concept of "magnification" is introduced so as to quantitatively evolve (to a finer scale) the originally static definition of resolution proposed by Archard (1953). In Persson model, the magnification is quantified by a parameter $\zeta$ as follows: at the lowest magnification rate, (i.e., $\zeta = 1$), no asperities are observed; when zooming in, $\zeta$ increases and asperities of smaller scales are gradually observed. Under the assumption of full contact, the probability density of the contact pressure is suggested to "diffuse" (in the contact pressure space), as the rough surface gets magnified. Then by imposing proper initial and boundary conditions, the actual contact area and contact force can be formulated. Benefitting from the idea of magnification, Persson model has found its value in many applications, e.g., leak-rate of sealing (Lorenz and Persson, 2010; Persson and Yang, 2008), influence of surface roughness on super-hydrophobicity (Yang et al., 2006a, 2008) and biological adhesion (Persson, 2008). It is worth noting that because of the full contact assumption employed so as to derive the related diffusion equation, Persson model is expected to deliver high accuracy at large indentation distance, and this has been verified by experimental (Lorenz et al., 2010) and numerical results (Almqvist et al., 2011; Yang and Persson, 2008b; Yastrebov et al., 2017). However, its accuracy becomes worse in the case of light indentations or small contact areas (Ciavarella, 2016; Papangelo et al. 2017), and such deviation is also seen in the comparison with experimental data (Lorenz et al., 2010).

Based on the above discussion, it is obvious that a theory which can be employed more directly and quantitatively to describe the contact status during the entire compression process (in consistency with the up-to-date identification approaches for rough surface) is still urgently needed for rough contacts without adhesion. This motivates the present study, where the idea of magnification introduced by Persson is effectively incorporated into the asperity-representation of rough surfaces presented by Greenwood and Williamson. The proposed magnification-based multi-asperity (MBMA) model is established based on the fact that the actual contact takes place on the top of a number of "contact islands", on which smaller-scale asperities are distributed. Then for a given indentation distance, only asperities whose sizes fall into a certain range play a crucial role in the determination of the corresponding contact status of each contact island. This is because the asperities of too small sizes have been flattened at this compression status, while the asperities of too large sizes have barely deformed. With this assumption solidly supported by the theoretical analysis (Majumdar and Bhushan, 1991) and experimental studies (Whitehouse and Archard, 1970), the proposed MBMA depicts a fairly clear physical picture, and the key point for solving the problem is to determine the effect range for the asperities curvature radii associated with each contact island, which should evolve along with the indentation distance. By doing this, the originally complex contact problem with fractal rough surfaces is equivalently transformed to a set of contact problems between an elastic sphere whose curvature radius equals the lower limit of the aforementioned effective range (associated with each contact island) and a new rough surface consisting of asperities whose curvature radii fall into the effective range associated with each contact island. Based on the idea of Greenwood and Tripp (1967), the contact information is then formulated for each contact island, and the overall contact status is calculated by summarizing over the contributions from all contact islands. As a validation of the derived MBMA model, it is well compared with the experimental (Lorenz, 2012; Lorenz et al., 2010; Putignano et al., 2013) and numerical (Papangelo et al., 2017; Yang and Persson, 2008a; Yang and Persson, 2008b; Yang et al., 2006b) examples recorded in literature. Especially, the simulation results based on the MBMA model are well compared with the representative theoretical, numerical and experimental data that are recently summarized in a "contact challenge" test held by Mueser et al. (2017). Another interesting finding is that the G-W approach (which works well for small deformation) and the Persson approach (which works well for deep indentation, i.e., large deformation) is well connected by the proposed MBMA theory. Furthermore, a detailed study has also been performed over the scale effect observed in contact forces. Again, comparing to the G-W and Persson models, the derived MBMA theory is shown to deliver relatively accurate predictions during the entire compression process.

Besides, the proposed MBMA model is also employed to investigate several key issues which have been intensively addressed when comparative analysis between multi-asperity contact models and Persson model is made (Bottiglione et al., 2009; Campaná and Mueser, 2007; Carbone and Bottiglione, 2008, 2011; Hyun et al., 2004; Lorenz et al., 2010; Persson, 2006; Putignano et al., 2012; Yastrebov et al., 2017). For example, in the limit of small compression, some studies (Campaná and Mueser, 2007; Carbone and Bottiglione, 2008; Ciavarella et al., 2006b; Hyun et al., 2004; Persson, 2006; Putignano et al., 2012; Putignano et al., 2013; Yastrebov et al., 2017) have been carried out to calculate the non-dimensional scaling coefficient arising in the linear relationship between the actual contact area and the contact force. With use of the derived MBMA model, this parameter is found to be almost 2, which is closer to the numerical predictions (finite element methods and molecular dynamic simulations) (Campaná and Mueser, 2007; Hyun et al., 2004; Putignano et al., 2012; Putignano et al., 2013), compared to those obtained from other multi-asperity contact models ($\sqrt{2\pi}$) and Persson model ($\sqrt{8/\pi}$) (Akarapu et al., 2011; Campaná and Mueser, 2007; Carbone and Bottiglione, 2008; Hyun et al., 2004; Persson, 2006; Putignano et al., 2012; Yastrebov et al., 2017). The present work also aims for addressing the issues on parameter transformation when the G-W (and alike) and Persson approaches are compared. For example, the separation between the rigid surface and the original mid-plane of the elastic rough surface used in the multi-asperity contact models, is naturally used as the average interfacial separation as proposed by Persson (Carbone and Bottiglione, 2011; Lorenz et al., 2010). However, the two quantities may be essentially different. For this purpose, a systematic study over the parameter transformation from the G-W framework and alike to that of Persson's is also carried out in this article.

The reminder of the paper is organized as follows. In Section 2, existing means for identifying a rough surface are summarized, along with a brief review of the G-W and Persson models. This is followed by the derivation of the MBMA model in Section 3. In Section 4, the proposed MBMA approach is validated through comparing with the theoretical, experimental and numerical results recorded in literature. The article concludes with several remarks in Section 5. Noted that in order to facilitate further reading of the present article, frequently used notations are listed in Appendix A.

## 2. Review of the theory of rough contact in literature

In this section, the characteristics of rough surfaces and their representation are introduced first. Then two most widely used theories for studying rough contact problems, the G-W approach and the Persson approach, are briefly reviewed.

### 2.1. Characteristics and representation of a rough surface

Given a longitudinal section of a rough surface along any transverse direction as shown in Fig. 1, the height of this profile is denoted by $z(x)$. Given a rough surface of profile $z(x)$, three statistical quantities, known as Nayak's moments are found effectively summarizing its microscopic information. They are the variances of the height, slope and second derivative of $z(x)$ shown as follows:

$$m_0 = E\langle z^2 \rangle, m_2 = E\left\langle \left(\frac{dz}{dx}\right)^2 \right\rangle, m_4 = E\left\langle \left(\frac{d^2z}{dx^2}\right)^2 \right\rangle, \tag{1}$$

where $E\langle \cdot \rangle$ denotes the operation of taking expectation. Once $m_0, m_2, m_4$ are known, according to the studies by Nayak (1971) and McCool (1986), the surface asperity density $D_{\text{sum}}$, the average asperity summit curvature radius $R$ and the standard deviation of the summit heights $\sigma$ for an isotropic surface can be evaluated by

$$D_{\text{sum}} = \frac{m_4}{6\pi\sqrt{3}m_2}, R = \frac{3\sqrt{\pi}}{8\sqrt{m_4}}, \sigma = \sqrt{\left(1 - \frac{0.8968}{\alpha}\right)m_0}, \tag{2}$$

respectively, where $\alpha$ is known as the Nayak's parameter satisfying

$$\alpha = m_0 m_4 / m_2^2. \tag{3}$$

The practical quantification of Nayak's moments with direct use of Eq. (1), however, is not so straightforward mainly due to two issues: i) a rough surface is a two-dimensional object while Nayak's moments are defined in a one-dimensional contact; ii) the profile of a rough surface normally displays a non-differentiable feature at a finest scale (as shown in Fig. 1). Alternatively, Nayak's moments are usually calculated by means of the measured power spectral density function $C(\mathbf{q})$, where $\mathbf{q} = (q_1 = 2\pi/\lambda_1, q_2 = 2\pi/\lambda_2)^T$ is a two-dimensional wavenumber vector with $\lambda_i, i = 1,2$ denoting the wave lengths along the two coordinate directions. The introduction of $C(\mathbf{q})$ is based on the postulate that the profile of a rough surface can be expressed by the superimposition of a family of sinusoidal waves (Greenwood and Wu, 2001; Persson, 2001; Wilson et al., 2010). In theory, $C(\mathbf{q})$ is defined by (Greenwood and Wu, 2001; Mohammadi et al., 2013; Persson, 2001, 2006; Whitehouse and Archard, 1970)

$$C(\mathbf{q}) = \frac{1}{4\pi^2} \iint_{-\infty}^{+\infty} R(\mathbf{L}) e^{-i\mathbf{q}\cdot\mathbf{L}} d\mathbf{L}, \tag{4}$$

where the auto-correlation function $R(\mathbf{L})$ satisfies

$$R(\mathbf{L}) = E\langle z(\mathbf{x})z(\mathbf{x} + \mathbf{L})\rangle. \tag{5}$$

In practice, $z(\mathbf{x})$ is only measured at a set of grid points, and $\mathbf{L}$ in Eqs. (4) and (5) is calculated as takes only the integer values multiplied by the sampling spacing. By doing this, $C(\mathbf{q})$ in Eq. (4) can be computed by using the fast Fourier transform (FFT) technique. It is worth noting that the experimental quantification of a rough surface in this way is assigned with a resolution, which is determined by the sampling spacing.

Moreover, under the assumption that a rough surface is an isotropic self-affine fractal (Borodich, 2013; Persson, 2014), the spectral density function further becomes a function of only the magnitude of the wavevector $q = |\mathbf{q}| = \sqrt{q_1^2 + q_2^2}$, and it can be expressed in terms of a power-law function (Bottiglione et al., 2009; Persson, 2001, 2006)

$$C(q) = \frac{H}{\pi} \frac{h_{\text{rms}}^2}{(q_0^{-2H} - q_1^{-2H})} q^{-2(1+H)} \qquad (6)$$

for $q_0 \leq q \leq q_1$. In Eq. (6), $H$ is the Hurst exponent; $h_{\text{rms}}$ quantifies the mean square root roughness of the surface; $q_0$ is the long distance roll-off wavenumber (related to the largest scale of the asperity); $q_1$ is the short distance cut-off wavenumber (related to the smallest scale of the asperity that is experimentally observable), which equals $\pi$ divided by the sampling spacing. Thus the interval of the characteristic wavenumbers $q$ is highly dependent on the sampling scheme adopted (as shown in Fig. 2).

It is noted that the power-law type power spectral density function given by Eq. (6) (also shown in Fig. 2) is employed in this article. When all parameters in $C(q)$ are determined, the Nayak's moments is then calculated by (Bottiglione et al., 2009; Bush et al., 1975; Lorenz et al., 2010; Yastrebov et al., 2017):

$$m_n = \int_0^{2\pi} (\cos\varphi)^n \mathrm{d}\varphi \int_{q_0}^{q_1} C(q) q^{1+n} \mathrm{d}q. \qquad (7)$$

A comparison between Eqs. (6) and (7) gives the expression for the mean square root roughness $h_{\text{rms}}$ (Almqvist et al., 2011; Yang and Persson, 2008a):

$$h_{\text{rms}} = \sqrt{m_0} = \left(2\pi \int_{q_0}^{q_1} C(q) q \mathrm{d}q\right)^{1/2}. \qquad (8)$$

Combining Eqs. (2), (6) and (7), the values of $D_{sum}$, $R$ and $\sigma$ of a rough surface can all be evaluated.

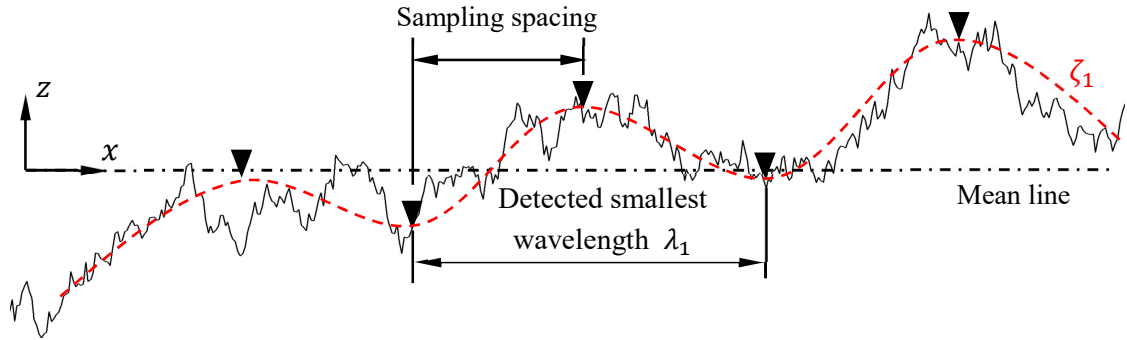

Fig.1. A cut through a fractal rough surface.

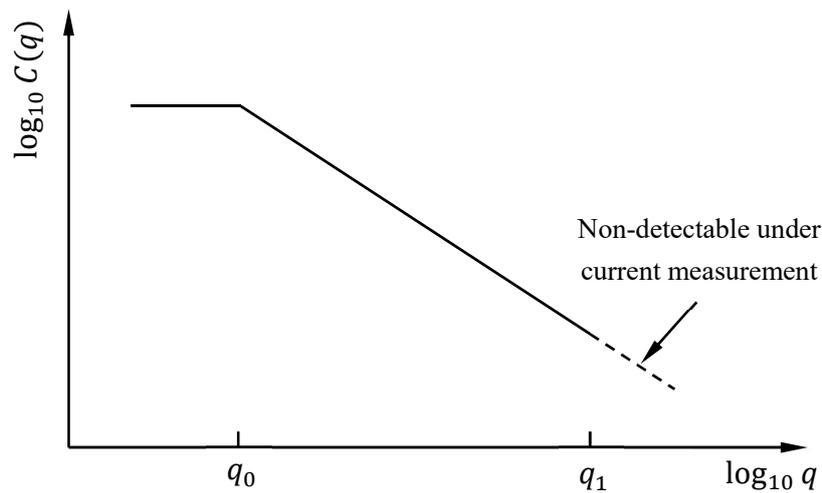

Fig.2. The power spectrum density function of a self-affine fractal rough surface (Pelliccione and Lu, 2008; Persson, 2006; Yang and Persson, 2008a).

## 2.2. A review of the G-W model

In the G-W framework (Greenwood and Williamson, 1966), the contact between two rough surfaces is modelled as a contact problem between a rigid flat surface and an elastic solid decorated with a randomly rough surface, which is further assumed to be an ensemble of non-interacting spherical asperities of the same curvature radius $R$ (see Fig. 3 for reference). Meanwhile, the asperity height $z$ measured from the mid-plane of the rough surface is assumed to satisfy a Gaussian distribution with the probability density function $\phi(z)$ defined as

$$\phi(z) = \frac{1}{\sqrt{2\pi}\sigma} \exp\left(-\frac{z^2}{\sigma^2}\right), \tag{9}$$

where $\sigma$ is recalled to be the standard deviation of the asperity heights defined in Eq. (2). When the rough surface is compressed such that the distance between the rigid flat surface and the corresponding mid-plane of the rough surface equals $d$ (see Fig. 3 for reference), the asperities whose heights are greater than $d$ should deform, and the corresponding probability is calculated by

$$\text{Prob}(z \geq d) = \int_d^\infty \phi(z)\,dz. \tag{10}$$

Given the surface summit density $D_{\text{sum}}$ obtained from Eq. (2), the actual contact area is then calculated as

$$A = A_0 D_{\text{sum}} \pi R \int_d^\infty (z-d)\phi(z)\,dz, \tag{11}$$

where $A_0$ is the nominal contact area. By assuming that the deformation of asperity satisfies Hertz's solution of plane-sphere contact (Hertz, 1881), the expected contact force is thus calculated by

$$F = \frac{4}{3} A_0 D_{\text{sum}} E^* R^{1/2} \int_d^\infty (z-d)^{3/2} \phi(z)\,dz, \tag{12}$$

where $E^* = [(1-\nu_1^2)/E_1 + (1-\nu_2^2)/E_2]^{-1}$ is the equivalent Young's modulus for contact analysis with $\nu_i$ and $E_i$, $i = 1,2$ denoting the Poisson ratios and Young's modulus of the two elastic solids, respectively (Johnson, 1987).

It is worth noting that the idea underlying the G-W's approach has also been generalized to establish models of rough surface with multiple representative shapes and curvature radii (Bush et al., 1975; Greenwood, 2006; Nayak, 1971). The rough surfaces in G-W framework and alike are characterized in a single-scaled sense (Carbone and Bottiglione, 2011; Thomas, 1998), i.e., the status at all contact points from the originally multiscale rough surface is formulated in an averaged sense. Such treatment may result in inaccuracy for deep indentation. This is because the well known Hertz's formulation is normally valid for small deformations. When the rough surface gets deeply compressed, however, small-scale asperities may have experienced large indentation, even been flattened (Whitehouse and Archard, 1970). Hence, G-W single-scaled description model and alike where Hertz formulation has been adopted are more preferred in the case when the deformation is small (Carbone and Bottiglione, 2008, 2011; Lorenz et al., 2010; Persson, 2006; Thomas, 1998).

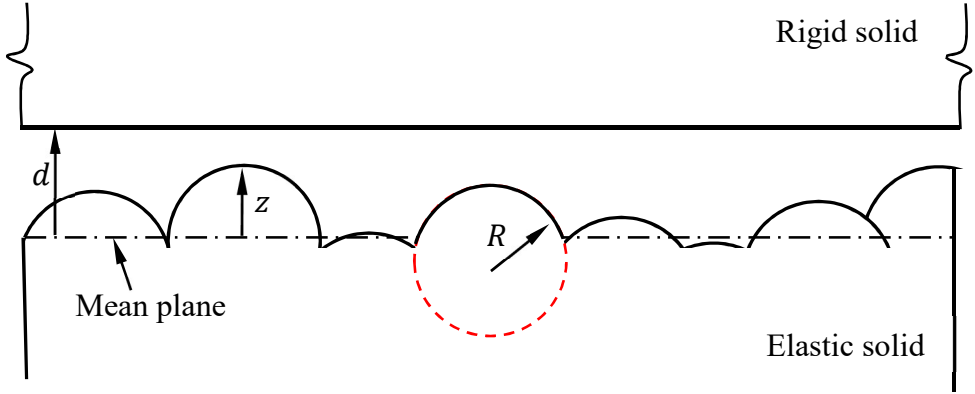

Fig. 3. The G-W model for rough contact

### 2.3. A review of the Persson model

In order to capture the multiscale of contact behavior between rough surfaces, Persson (2001) introduced the concept of magnification in his rough contact theory. In Persson model, the originally complex multiscale contact problem is transformed to a problem where a rigid rough surface is compressed by an elastic flat surface. Given a wavenumber $q$, a (non-dimensional) magnification parameter $\zeta$ is defined by (Persson, 2001, 2006)

$$\zeta = \frac{q}{q_0}, \tag{13}$$

where $q_0$ is recalled to be the smallest wavenumber. Given a magnification $\zeta$, only asperities whose wavenumbers fall below $q$ are observable. When $\zeta = 1$, as shown in Fig. 4, no asperities are captured, and this corresponds to the case where the two contacting surfaces are both flat (Persson, 2001, 2006; Bottiglione et al., 2009). The magnification hits its upper limit at $\zeta_1 = q_1/q_0$, where $q_1$ is recalled to be the maximum wavenumber that can be measured experimentally.

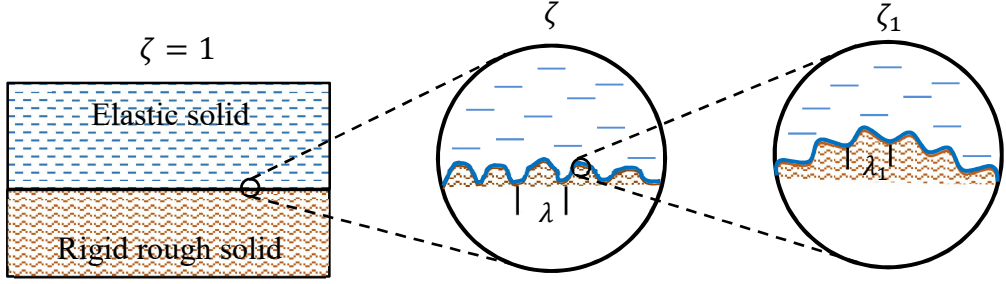

Fig 4. The full contact postulate observed at different magnifications $\zeta$.

In Persson model, full contact status at any magnification is first assumed as shown in Fig. 4 (i.e., the elastic solid fills all the cavities between the contact surfaces) and a probability density function $P(p,\zeta)$ is assigned to each magnification $\zeta$, where $p$ is the contact stress (Persson, 2001). As the rough surface gets magnified, this probability density $P$ is assumed to "diffuse" in the space of the contact stress:

$$\frac{\partial P}{\partial \zeta} = \frac{\mathrm{d}G}{\mathrm{d}\zeta} \cdot \frac{\partial^2 P}{\partial p^2}, \tag{14}$$

where $G$ carries the information of the rough surface. Under the postulation of full contact, $G$ can be expressed by $G(\zeta) = E^{*2} m_2(\zeta)/4$, where $m_2(\zeta)$ is calculated in a similar way as in Eq. (7) except for the upper limit (because of the magnification):

$$m_2(\zeta) = \pi \int_{q_0}^{\zeta q_0} C(q) q^3 \mathrm{d}q. \tag{15}$$

As for the lowest magnification $\zeta = 1$, the problem is equivalent to the contact problem between two flat surfaces, and this results in the initial condition:

$$P(p, 1) = \delta(p - p_0), \tag{16}$$

where $\delta(\cdot)$ is the Dirac delta function; $p_0$ is the nominal contact stress given by $p_0 = F/A_0$ with $F$ and $A_0$ being the total contact force and the nominal contact area, respectively (Persson, 2001, 2006). Under the full contact assumption, Eq. (14) holds for $(p, \zeta) \in (-\infty, +\infty) \times [1, \zeta_1]$, where $p < 0$ corresponds to adhesive interaction. However, since the actual contact is only made without adhesion, a boundary condition

$$P(0, \zeta) = 0 \tag{17}$$

is imposed, then the expression for $P(p, \zeta)$ with $p > 0$ can be obtained by solving

Eq. (14) along with Eqs. (16) and (17) (Manners and Greenwood, 2006). It is noted that $P(p,\zeta)$ satisfying Eq. (17) is no longer a probability density function for $p \in (0,+\infty)$, and its integration with respect to $p$ over $(0,+\infty)$ gives the ratio of the actual contact area to the nominal area (which depends on the magnification $\zeta$ and the nominal contact stress $p_0$):

$$\frac{A(\zeta;p_0)}{A_0} = \int_0^{+\infty} P(p,\zeta)\,dp = \mathrm{erf}\left(\frac{p_0}{2\sqrt{G}}\right), \tag{18}$$

where $\mathrm{erf}(\cdot)$ is the error function. In addition, the average interfacial separation $u$ between the two contact rough surfaces can be calculated as (Persson, 2007; Yang and Persson, 2008a)

$$u(p_0) = \sqrt{\pi}\int_{q_0}^{q_1} q^2 C(q) w(q) \times \int_{p_0}^{\infty} \frac{\gamma + 3(1-\gamma)\left[\frac{A(\zeta;p')}{A_0}\right]^2}{p'} e^{-\left[\frac{w(q)p'}{E^*}\right]^2} dp'\,dq, \tag{19}$$

where $\gamma = 0.42$ is a constant; $w$ is related to $G$ by means of the equivalent Young's modulus $E^*$ as $w = E^*/2\sqrt{G}$. Given the complication in the double integral appearing in Eq. (19), a simpler expression for $u$ is suggested as (Yang and Persson, 2008a):

$$u = -u_0 \ln\left(\frac{F}{E^*\lambda A_0}\right), \tag{20}$$

where

$$u_0 = \sqrt{\pi}\gamma \int_{q_0}^{q_1} q^2 C(q) w(q)\,dq \tag{21a}$$

and

$$\lambda = 4.047 \times \exp\left(-\frac{\int_{q_0}^{q_1} q^2 C(q) w(q) \ln[w(q)]\,dq}{\int_{q_0}^{q_1} q^2 C(q) w(q)\,dq}\right), \tag{21b}$$

respectively. As will be discussed in Section 4, compared to G-W model, Persson model delivers more accurate results for deep indentation. Nevertheless, owing to its full contact postulate, the problem may become less accurate for small contact areas and loads.

# 3. A magnification-based multi-asperity model

## 3.1. General strategy

As summarized above, the magnification parameter $\zeta$ introduced by Persson helps to capture the multiscale feature displayed by rough surfaces, while the identification of rough surfaces with representative asperities in G-W model provides a simple but effective description of the actual contact status (for small deformation). In this section, by integrating the aforementioned features displayed by both G-W and Persson models, we shall derive a magnification-based multi-asperity (MBMA) model for predicting the contact status during the entire compression process.

As shown in Fig. 5, at low magnification, a rough surface can be equivalently considered as a number of asperities of large curvature radii (say $R_d^i$ in Fig. 5), since smaller-scale asperities are non-observable. At high magnification, the contact on smaller-scale asperities (with curvature radii $R_u^i$ if Fig. 5 is referred to) is resolved. From this viewpoint, a rough surface can be treated as a series of asperities where the asperities at each level are decorated with smaller-size asperities. Such an asperity series can be schematically represented by a tree structure shown in Fig. 6. The root of the tree corresponds to the profile observed when $\zeta = 1$ (see Fig. 4 for reference). As the profile gets magnified, smaller asperities which are decorated on originally large-asperities emerge. Correspondingly, branches representing asperities of smaller scales grow from a node representing an asperity of large radius. It is physically meaningful to say that, when being compressed, asperities with too small curvature radii may have already been flattened, while asperities with too large curvature radii have barely deformed. In addition, the tree structure shown in Fig. 6 indicates that only asperities falling in the shaded region play a major role in the formulation of the actual contact force. This means that a rough surface can be further represented by a set of subtrees as highlighted in Fig. 6. The root of each sub-tree corresponds to a contact island depicted in Fig. 5. Therefore, if we manage to calculate the indentation distance assigned to each contact island, we can estimate an interval for the curvature radii of asperities $[R_u^i, R_d^i]$ which respond effectively to the indentation assigned to the $i$-th subtree, and the contact status of the $i$-th island can be formulated individually. The overall contact behavior on the rough surface thus can be formulated by summarizing the contributions from individual subtrees.

Based on the aforementioned analysis, the strategy of establishing the proposed MBMA model is summarized as follows.

1) *Determination of the effective range* $[R_\mathrm{u}^i, R_\mathrm{d}^i]$ *for each sub-tree*. For the $i$-th contact island, $R_\mathrm{d}^i$ equals the radius of the contact island which corresponds to the asperity located at the root of the $i$-th sub-tree. $R_\mathrm{u}^i$ equals the shortest curvature radius of all asperities that still respond effectively to the indentation assigned to the $i$-th contact island.

2) *Determination of the contact status associated with the $i$-th contact island*. This is achieved by employing the formulation proposed by Greenwood and Tripp (1967), where the contact status between an elastic sphere and an elastic rough surface is studied.

3) *Determination of the overall contact status of the rough surface*. This is obtained by summarizing over the contributions from all contact islands.

It is worth noting that the interval $[R_\mathrm{u}^i, R_\mathrm{d}^i]$ (the shaded region in Fig. 6) should evolve as the indentation proceeds. This is because a deeper indentation may induce considerable deformation to asperities that have barely deformed originally.

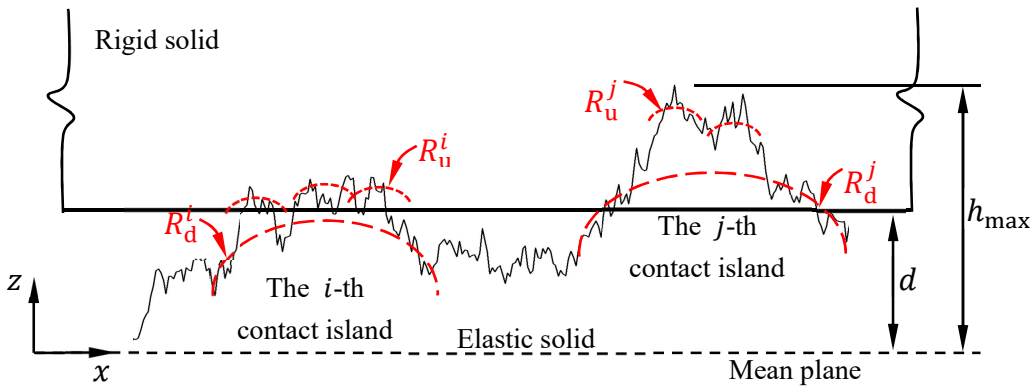

Fig. 5. A rough surface can be treated as an ensemble of spherical asperities with different sizes: at low magnification, it can be considered as a number of asperities of large curvature radii (say $R_\mathrm{d}^i$ or $R_\mathrm{d}^j$); while at high magnification, it can be considered as a number of asperities of small curvature radii (say $R_\mathrm{u}^i$ or $R_\mathrm{u}^j$).

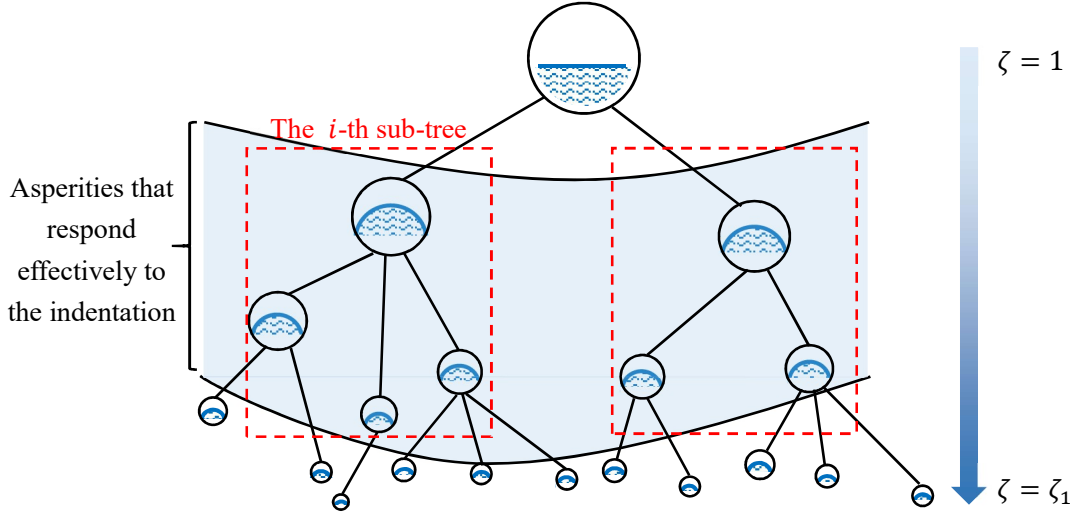

Fig. 6. The tree structure representing the asperity series of a rough surface: the root of each sub-tree corresponds to a contact island.

### 3.2. Determination of $R_\mathrm{u}^i$ and $R_\mathrm{d}^i$

#### 3.2.1. Relating the curvature radii of asperities to magnification $\zeta$

To detect the two bounds for the curvature radii of asperities in the $i$-th contact island (i.e., $R_\mathrm{u}^i$ and $R_\mathrm{d}^i$), we incorporate the idea of magnification into the G-W model. Under a given magnification $\zeta$ or equivalently the corresponding wavenumber $q$ (because $\zeta = q/q_0$), a rough surface can be considered as a family of identical spheres of radius $R'$ as suggested by Greenwood and Williamson (1966). To determine the effective radius $R'$, one can simply replace $q_1$ by $\zeta q_0$ in Eq. (2):

$$R' = \frac{3\sqrt{\pi}}{8\sqrt{m_4(\zeta)}}, \qquad (22)$$

where $m_n(\zeta)$ is the $n$-th Nayak's moments observed at magnification $\zeta$, which in analogy to Eq. (7), are given by

$$m_n(\zeta) = \int_0^{2\pi} (\cos\varphi)^n \mathrm{d}\varphi \int_{q_0}^{\zeta q_0} C(q) q^{1+n} \mathrm{d}q. \qquad (23)$$

Hence with use of Eqs. (22), (23) and $\zeta = q/q_0$, the determination of $[R_u^i, R_d^i]$ is equivalent to the calculation of an interval of magnifications $[\zeta_d^i, \zeta_u^i]$. It is noted that $R_u^i < R_d^i$, but $\zeta_u^i > \zeta_d^i$. Under this setup, all asperities that can already be observed at magnification $\zeta_d^i$ are considered to be too rigid to deform against indentation, while all asperities that cannot be observed even at magnification $\zeta_u^i$ should be treated as being flattened.

In order to determine $\zeta_d^i$ and $\zeta_u^i$, here we refer to the works by Greenwood and Tripp (1967) as shown in Fig. 7. A rough surface with the mean square root roughness $\sigma$ compressed by a rigid plane can be equivalently as a contact problem between two rough surfaces (whose mean square root roughness are $\sigma_1$ and $\sigma_2$, respectively), where $\sigma^2 = \sigma_1^2 + \sigma_2^2$ (Johnson, 1987). To be specific, for any wavenumber $q \in [q_0, q_1]$, the two rough surfaces shown in Fig. 7(b) can be determined as follows. One rough surface is an elastic rough surface (the upper one) consisting of (small) asperities whose wavenumbers fall in $[q, q_1]$. The other one (the lower one) can be treated as elastic spherical asperities of identical curvature radius through averaging over all asperities whose wavenumbers fall in $[q_0, q]$.

The quantities that will be of frequent use in the rest of the calculation are the mean square root roughness associated with the two elastic surfaces in Fig. 7(b). Here we quantify the mean square root roughness of the upper elastic rough surface in Fig. 7(b) by $\overline{h}$ and the lower elastic surface consisting of spherical asperities in Fig. 7(b) by $\underline{h}$. To calculate $\underline{h}$, one refers to the expression for the mean square root roughness of the original rough surface $h_{\text{rms}}$ in Eq. (8). Replacing the upper limit of the integral in Eq. (8) by $q_0\zeta$ gives

$$\underline{h} = \left(2\pi \int_{q_0}^{\zeta q_0} C(q)q\mathrm{d}q\right)^{1/2} = h_{\text{rms}}\sqrt{\frac{1 - \zeta^{-2H}}{1 - \zeta_1^{-2H}}}. \tag{24}$$

Correspondingly, the mean square root roughness of the upper elastic rough surface in Fig. 7(b) is quantified by

$$\overline{h} = \left(2\pi \int_{\zeta q_0}^{q_1} C(q)q\mathrm{d}q\right)^{1/2} = h_{\text{rms}}\sqrt{\frac{\zeta^{-2H} - \zeta_1^{-2H}}{1 - \zeta_1^{-2H}}}. \tag{25}$$

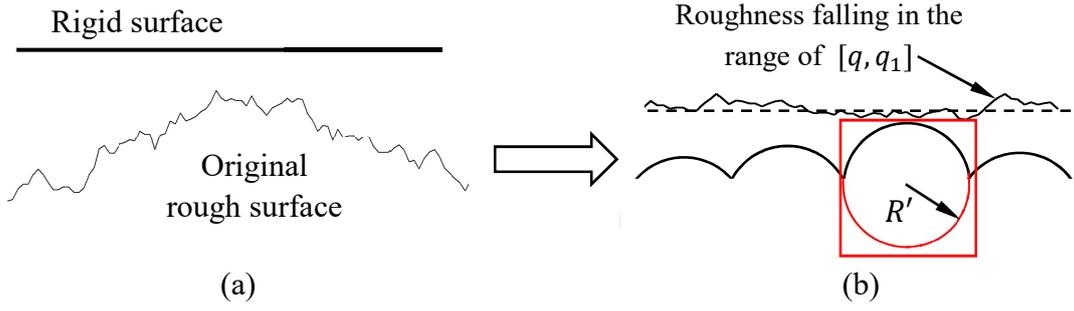

Fig. 7. To determine $R_d^i$ and $R_u^i$, the contact problem can be viewed as being equivalent to the contact between a rough surface (constructed based on the roughness in the range of $[q, q_1]$) and equivalent sphere asperities with the same radius $R'$ (created by the roughness in the range of $[q_0, q]$).

### 3.2.2. Estimation of $R_d^i$ and $R_u^i$

Now we calculate the largest curvature radius $R_d^i$ of asperities that still respond effectively to an indentation distance assigned to the $i$-th contact island $\delta^i = z_{\max}^i - d$ (as shown in Fig. 8). Based on the analysis in Section 3.1, it is noted that $R_d^i$ equals the radius of the contact island which corresponds to the asperity located at the root of the $i$-th sub-tree (see Fig. 6 for reference). To determine $R_d^i$, we need to identify the critical wavenumber $\zeta_d^i$, under which the corresponding asperities may have barely deformed. The translation depicted in Fig. 7 is now referred to. In this scenario, the upper elastic rough surface in Fig. 7(b) should consist of asperities whose characteristics wavenumbers fall in the range of $[q_0 \zeta_d^i, q_1]$. On the other hand, the lower surface in Fig. 7(b) should consist of asperities of radius $R_d^i$. To calculate the value of $\zeta_d^i$ and $R_d^i$, we refer to the results by Phort and Popov (2013), where the contact status between an elastic rough sphere and a rigid plane is investigated by using three-dimensional boundary element methods. They pointed out that when the stiffness of an elastic sphere is greater than that induced by its surface roughness, it is the roughness of the sphere that dominates the deformation. We refer the readers to Appendix B for more technical details, and here the formulations for $R_d^i$ are directly listed as:

$$R_{\mathrm{d}}^i = \frac{3\sqrt{\pi}}{8\sqrt{m_4(\zeta_{\mathrm{d}}^i)}}, \tag{26}$$

where $m_4(\zeta_{\mathrm{d}}^i)$ is evaluated based on Eq. (23). Meanwhile, $\zeta_{\mathrm{d}}^i$ satisfies

$$\zeta_{\mathrm{d}}^i = \max\left(1.5, \underline{\zeta}^i\right) \tag{27a}$$

and

$$\underline{\zeta}^i = \left\{ \left[\frac{4}{9}\left(\frac{z_{\max}^i - d}{h_{\mathrm{rms}}}\right)\right]^2 + \zeta_1^{-2H} \right\}^{-1/2H}. \tag{27b}$$

To determine $R_{\mathrm{u}}^i$, we need to identify the critical wavenumber $\zeta_{\mathrm{u}}^i$, over which the corresponding asperities become negligible (as shown in Fig. 8). Again, we refer to the translation depicted in Fig. 7. Correspondingly, the radius of the elastic spheres (in Fig. 7(b)) should equal $R_{\mathrm{u}}^i$, which is obtained by averaging the curvature radii of all asperities from $[q_0, q_0\zeta_{\mathrm{u}}^i]$. Then the elastic rough surface (in Fig. 7(b)) should only consist of asperities corresponding to wavenumbers falling in $[q_0\zeta_{\mathrm{u}}^i, q_1]$. To calculate the critical wavenumber $\zeta_{\mathrm{u}}^i$, we refer to the conclusion by Greenwood and Tripp (1967) and Johnson (1987), where a non-dimensional parameter $\varphi$ is introduced, such that

$$\varphi = \frac{\sigma_{\mathrm{u}1}}{\delta_{\mathrm{u}}}, \tag{28}$$

where $\sigma_{\mathrm{u}1}$ is the standard deviation of the summit heights of the new elastic rough surface (in Fig. 7(b)); $\delta_{\mathrm{u}}$ is the Hertz deformation of the asperities of radius $R_{\mathrm{u}}^i$ (Johnson, 1987). It is found that when $\varphi \leq 0.05$, the effect due to the presence of the rough surface is negligible (Bahrami et al., 2005; Greenwood et al., 1984). We refer the readers to Appendix B for a detailed derivation, and here the formulations for $R_{\mathrm{u}}^i$ are directly listed as:

$$R_{\mathrm{u}}^i = \frac{3\sqrt{\pi}}{8\sqrt{m_4(\zeta_{\mathrm{u}}^i)}}, \tag{29}$$

where $m_4(\zeta_{\mathrm{u}}^i)$ is evaluated based on Eq. (23). Meanwhile, $\zeta_{\mathrm{u}}^i$ satisfies

$$\zeta_{\mathrm{u}}^i = \max\left(3, \overline{\zeta}^i\right) \tag{30a}$$

and

$$\bar{\zeta}^i = \left[\left(\frac{z_{\max}^i - d}{21}\right)^2 \cdot \frac{1-\zeta_1^{-2H}}{h_{\text{rms}}^2} + \zeta_1^{-2H}\right]^{-\frac{1}{2H}}. \tag{30b}$$

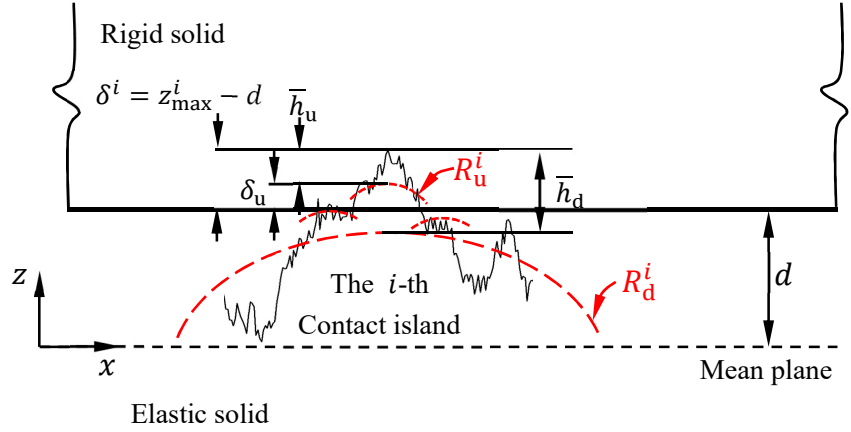

Fig. 8. Contact analysis at the $i$-th contact island under an indentation distance $\delta^i$.

### 3.3. Determination of the contact status of the $i$-th contact island

Once $R_{\text{u}}^i$ and $R_{\text{d}}^i$ determined, we then further investigate the contact status of the $i$-th contact island (or choosing the $i$-th subtree shown in Fig. 6) with the total indentation quantified by $\delta^i = z_{\max}^i - d$ shown in Fig. 8. Now the rough surface of interest only consists of asperities whose wavenumbers fall in the range $[q_0\zeta_{\text{d}}^i, q_0\zeta_{\text{u}}^i]$. Therefore, the discretization treatment of the rough surface introduced by Greenwood and Tripp (1967) and Johnson (1987) is employed to translate the original problem to a set of new problems as shown in Fig. 9. In each of the new problem (corresponding to the $i$-th contact island), the contact takes place between an elastic spherical asperity of radius $R_{\text{d}}^i$ and an elastic rough surface consisting of asperities with wavenumbers belonging to $[q_0\zeta_{\text{d}}^i, q_0\zeta_{\text{u}}^i]$. With reference to the quantities defined in Section 2.1 for identifying a rough surface, we list the expressions for the characteristic quantities associated with the $i$-th new rough surface (see Fig. 9 for reference) as follows: the Nayak's moments are given by

$$m_n^i(\zeta) = \int_0^{2\pi} (\cos\varphi)^n \, d\varphi \int_{\zeta_{\text{d}}^i q_0}^{\zeta_{\text{u}}^i q_0} C(q) q^{1+n} \, dq, \tag{31}$$

the surface asperity density $D_{\text{sum}}^i$, the standard deviation of the summit heights $\sigma^i$ and the root mean square roughness $h_{\text{rms}}^i$ of the new rough surface are calculated by

$$D_{\text{sum}}^i = \frac{m_4^i}{6\pi\sqrt{3}m_2^i}, \sigma^i = \sqrt{\left(1 - \frac{0.8968}{\alpha^i}\right)m_0^i}, h_{\text{rms}}^i = \sqrt{m_0^i}, \quad (32)$$

respectively. In Eq. (32), $\alpha^i$ satisfies

$$\alpha^i = m_0^i m_4^i / (m_2^i)^2. \quad (33)$$

In addition, the datum plane is taken as the mid-plane of the rough surface. If the z-axis is set pointing vertically to the rough surface from the lowest point of the elastic sphere shown in Fig. 9, all quantities can be expressed as a function of $r$, the distance to the z-axis, where the postulation of isotropic rough surface is employed again. Thus the separation $d_1$ between the deformed sphere surface and the mid-plane of the rough surface can be calculated as (Greenwood and Tripp, 1967; Johnson, 1987)

$$d_1(r) = w_{\text{down}}(r) - w_{\text{down}}(0) + d_1(0) + r^2/2R_d^i, \quad (34)$$

where $w_{\text{down}}(r)$ is the deformation of the sphere at radius $r$. According to the schematic illustration shown in Fig. 9, the relationship

$$w_{\text{down}}(0) - d_1(0) = \delta^i - \bar{h}_d \quad (35)$$

Holds. In Eq. (35), $\bar{h}_d$ (calculated by Eq. (B.3)) is recalled to be the mean square root roughness falling in the range $[q_0\zeta_d^i, q_1]$ of the original surface. If the asperity deformation is elastic, the effective pressure at radius $r$ is given by (Greenwood and Tripp, 1967; Johnson, 1987)

$$p(r) = \frac{4}{3}\int_{d_1}^{\infty} E^* D_{\text{sum}}^i (R_u^i)^{1/2} [z_1 - d_1(r)]^{3/2} \frac{1}{\sqrt{2\pi}\sigma^i} \exp\left(-\frac{z_1^2}{2\sigma^{i2}}\right) dz_1. \quad (36)$$

Then the deformation of the sphere $w_{\text{down}}(r)$ can be related to the effective pressure $p(r)$ by the equations for the axi-symmetric deformation of an elastic half-space (Greenwood and Tripp, 1967; Johnson, 1987):

$$w_{\text{down}}(r) = \frac{4}{\pi E^*}\int_0^{\infty} \frac{t}{t+r} p(t) \boldsymbol{K}(k) dt, \quad (37)$$

where $\boldsymbol{K} = \boldsymbol{K}(k)$ is the elliptical integral of the first kind with the argument $k = 2(rt)^{1/2}/(r+t)$.

After the iterative computation of Eqs. (34), (36) and (37), the three unknowns in the equation system listed above: the sphere-roughness distance $d_1$, the effective pressure $p$ and the sphere deformation $w_{\text{down}}$ can all be obtained. In this scenario, the ratio of actual contact area to the nominal one in a small region at $r$ can be calculated by

$$a(r) = \pi\eta D_{\text{sum}}^i R_u^i \int_{d_1}^{\infty} [z_1 - d_1(r)] \frac{1}{\sqrt{2\pi}\sigma^i} \exp\left(-\frac{z_1^2}{2\sigma^{i2}}\right) dz_1, \quad (38)$$

where $\eta$ is defined as the ratio of the real contact area (when consider the ignoring roughness in the range $[q_0\zeta_u^i, q_1]$) to its Hertz contact area on the spherical asperities with the curvature radius $R_u^i$ (see the red box in Fig. 9). Actually, ignoring this part of roughness does not influence the calculated contact force supported by the spherical asperities as shown in Fig. 9, but can significantly influence its contact area. It is known that this spherical asperity is mainly propped up by those ignored asperities. Based on the theory of Greenwood and Tripp (1967), the forces supported by this roughness can be converted into a continuum form of the contact stress (multiplying the forces by the asperity density of this ignoring roughness as shown in Eq. (C.7)) on the summit of each spherical asperity with the radius $R_u^i$. It is found that the converted contact stress is very close to the classical Hertz solution, i.e., the roughness lying in the range $[q_0\zeta_u^i, q_1]$ can be ignored from a mechanical view point. However, the real contact area should be calculated by summarizing the results over this ignoring roughness. Therefore, based on the analysis in appendix C, $\eta$ can be written by

$$\eta = \frac{3}{4}\pi^{1/3} D_{\text{sum-t}}^{1/3} \left(\frac{3R_{\text{top}}}{2}\right)^{\frac{2}{3}} \left(\frac{20\bar{h}_u}{R_u^i}\right)^{\frac{1}{3}}, \quad (39)$$

where $D_{\text{sum-t}}$ and $R_{\text{top}}$ are the surface asperity density and the average summit curvature radius relating to the magnification $\zeta_1$, which can be obtained from Eqs. (2), (6) and (7); $\bar{h}_u$ is recalled to be the mean square-root roughness of the roughens whose wavenumbers falling into the range of $[q_0\zeta_u^i, q_1]$, and its value can be obtained by Eq. (B.12). Therefore, once $p(r)$ and $a(r)$ are computed, the actual contact area and the load on the $i$-th contact island can be calculated as

$$F^i = \int_0^\infty 2\pi r p(r)\, \mathrm{d}r \qquad (40a)$$

and

$$A^i = \int_0^\infty 2\pi r a(r)\, \mathrm{d}r, \qquad (40b)$$

respectively.

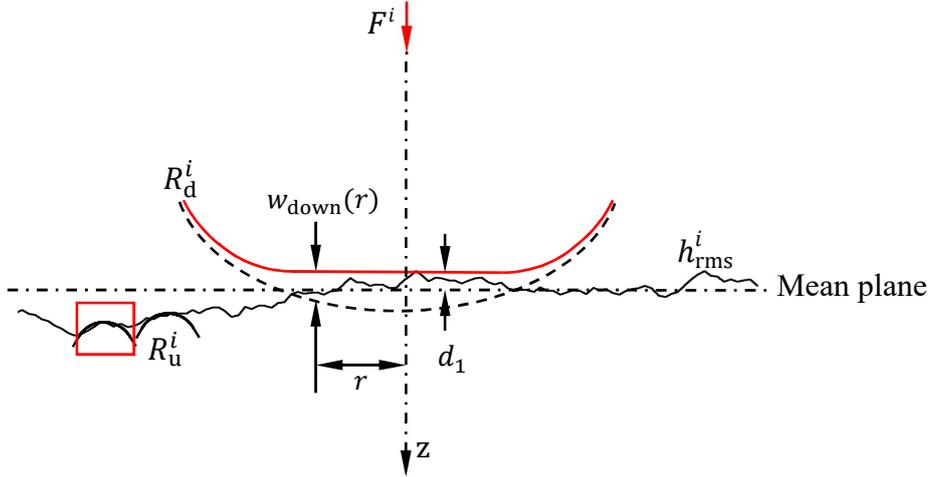

Fig. 9. A schematic illustration of the simplified theoretical model for the $i$-th contact island.

### 3.4. Determination of the overall contact status

Once the contact status ($A^i$ and $F^i$) for the $i$-th contact island is obtained, we further calculate the overall contact status, such as the actual contact area and the resulting contact force, by summarizing the results from individual contact islands. It is known that the $i$-th contact island is supported by a large spherical asperity observed at the magnification $\zeta_d^i$, and the height of this asperity denoted by $z'$ can be calculated by

$$z' = z_{\max}^i - \bar{h}_d, \qquad (41)$$

where $z_{\max}^i$ is recalled to be highest asperity height observable at the magnification $\zeta_1$ in the $i$-th contact island; $\bar{h}_d$ (which can be calculated from Eq. (B.3)) is recalled to be the mean square root roughness whose characteristic wavenumbers fall in the range $[q_0 \zeta_d^i, q_1]$ of the original surface.

We refer the readers to Fig. 8 for a schematic illustration of the above definitions.

Then the number of the contact islands with the same contact status on the contact surfaces is

$$N = A_0 D_{\text{sum}}^{\text{d}} \frac{1}{\sqrt{2\pi}\sigma_{\text{d}}} \exp\left(-\frac{(z')^2}{2\sigma_{\text{d}}^2}\right) dz', \quad (42)$$

where $D_{\text{sum}}^{\text{d}}$ and $\sigma_{\text{d}}$ is the density and the height standard deviation of the contact island, which corresponds to the spherical asperity located at the root of the $i$-th sub-tree with the radius $R_{\text{d}}^i$, respectively.

Actually, $D_{\text{sum}}^{\text{d}}$ and $\sigma_{\text{d}}$ are determined by the roughness whose characteristic wavenumbers in the range $[q_0, q_0\zeta_{\text{d}}^i]$, so they are obtained as

$$D_{\text{sum}}^{\text{d}} = \frac{m_4(\zeta_{\text{d}}^i)}{6\pi\sqrt{3}m_2(\zeta_{\text{d}}^i)}, \sigma_{\text{d}} = \sqrt{\left(1 - \frac{0.8968}{\alpha}\right)m_0(\zeta_{\text{d}}^i)} \quad (43a)$$

and

$$\alpha = \frac{m_0 m_4}{m_2^2}, m_n(\zeta) = \int_0^{2\pi} (\cos\varphi)^n d\varphi \int_{q_0}^{\zeta q_0} C(q) q^{1+n} dq, \quad (43b)$$

respectively.

Therefore, the total contact force on the contact surfaces can be calculated as

$$F = A_0 \int_{d-\overline{h}_{\text{d}}}^{h_{\max}-\overline{h}_{\text{d}}} F^i D_{\text{sum}}^{\text{d}} \frac{1}{\sqrt{2\pi}\sigma_{\text{d}}} \exp\left(-\frac{(z')^2}{2\sigma_{\text{d}}^2}\right) dz'. \quad (44)$$

Meanwhile, the real contact area is

$$A = A_0 \int_{d-\overline{h}_{\text{d}}}^{h_{\max}-\overline{h}_{\text{d}}} A^i D_{\text{sum}}^{\text{d}} \frac{1}{\sqrt{2\pi}\sigma_{\text{d}}} \exp\left(-\frac{(z')^2}{2\sigma_{\text{d}}^2}\right) dz'. \quad (45)$$

where $A^i$ and $F^i$ are the contact area and the load in the $i$-th contact island, respectively. Their values are obtained according to Eq. (40a) and (40b). Based on the relationship shown in Eq. (41), the integration variable in Eqs. (44) and (45) can be changed from $z'$ to $z_{\max}^i$ according to

$$dz' = \left(1 - \frac{d\overline{h}_{\text{d}}}{dz_{\max}^i}\right) dz_{\max}^i = K_1 dz_{\max}^i. \quad (46)$$

Combining Eqs. (B.3), (B.7), (B.10) and (46), the value of $K_1$ is obtained as

$$K_1 = \begin{cases} \dfrac{5}{9}, & \zeta_d^i > 1.5, \\ 1, & \zeta_d^i = 1.5. \end{cases} \quad (47)$$

At last, the total contact force $F$ and real contact area $A$ are rewritten as

$$F = A_0 \int_d^{h_{\max}} F^i D_{\text{sum}}^d \frac{1}{\sqrt{2\pi}\sigma_d} \exp\left(-\frac{(z_{\max}^i - \bar{h}_d)^2}{2\sigma_d^2}\right) K_1 dz_{\max}^i \quad (48)$$

and

$$A = A_0 \int_d^{h_{\max}} A^i D_{\text{sum}}^d \frac{1}{\sqrt{2\pi}\sigma_d} \exp\left(-\frac{(z_{\max}^i - \bar{h}_d)^2}{2\sigma_d^2}\right) K_1 dz_{\max}^i. \quad (49)$$

### 3.5. Numerical implementation of the MBMA approach

Based on the above analysis, a flow chart illustrating the implementation process of the proposed MBMA approach is shown in Fig. 10. Initially, the power spectral density $C(q)$ of the rough surface (characterized by $q_0$, $q_1$ and $H$), its mean square root roughness $h_{\text{rms}}$ and highest height $h_{\max}$ are measured by experimentation. Meanwhile, the current contact distance $d$ between the rigid flat surface and the mid-plane of the rough surface is also needed as an initial input. Then the two bounds for the curvature radii of asperities associated with the $i$-th contact island, i.e., $R_d^i$ and $R_u^i$ are firstly determined according to Eq. (26) and Eq. (29) in Section 3.2, respectively. Then by Eqs. (34), (36) and (37) in Section 3.3, the sphere-roughness distance $d_1$, the effective pressure $p$ and the sphere deformation $w_{\text{down}}$ can be obtained. Meanwhile, the contact force and contact area associated with the $i$-th contact island are calculated according to Eqs. (38), (40a) and (40b). Finally, as shown in Section 3.4, the overall contact force and contact area are determined based on Eqs. (48) and (49), respectively.

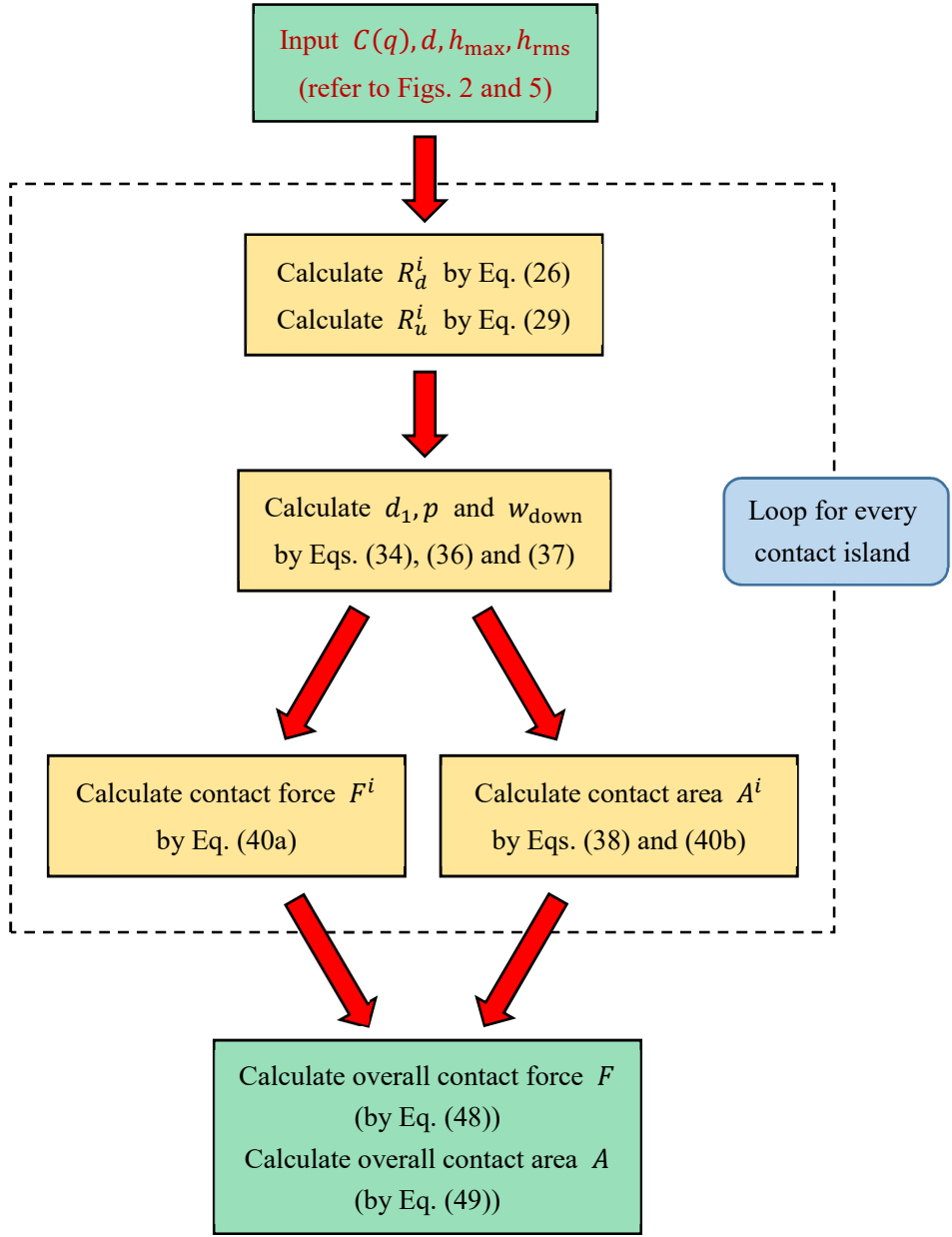

Fig. 10. Flow chart for the numerical implementation of the proposed MBMA model.

## 4. Numerical results and discussions

In this section, the effectiveness of the proposed MBMA model is illustrated by comparing its predictions with the available theoretical, experimental and numerical results in literature.

## 4.1. Comparison with theoretical/simulation results

### 4.1.1. Molecular dynamics simulation results by Yang and Persson

The first set of comparison concerns the molecular dynamics (MD) simulation results from Yang and Persson who studied the relationship between the contact force and the average interfacial separation (Yang and Persson, 2008a; Yang and Persson, 2008b; Yang et al., 2006b). The morphological parameters they used to generate the rough surface are listed as follow: the Hurst exponent $H = 0.8$; the long distance roll-off wavenumber $q_0 = 6\pi/1040 \times 10^{10} \text{m}^{-1}$; the mean square root roughness $h_{\text{rms}} = 10^{-9}\text{m}$; the largest measured wavenumber $q_1 = 216 q_0$; the highest height of the asperities $h_{\max} \approx 3 h_{\text{rms}}$. In Fig. 11, the computed contact forces are drawn against the average interfacial separation $u$ by using G-W model, Persson and Persson simplified model and the proposed MBMA model. The corresponding outcomes are compared with the MD simulation results. It is noted that in Fig. 11 and later, Persson's measure for the distance between two rough surfaces is used. In this case, the actual contact starts at $u = h_{\max} = 3 h_{\text{rms}}$, and it becomes fully contacted when $u = 0$.

Compared to the GW and two Persson models, the proposed MBMA model delivers an overall good match with the MD results, especially when the indentation is not deep, i.e., $u$ is large. It is noted that when $u$ exceeds $h_{\max}$ ($\approx 3 h_{\text{rms}}$), the two surfaces are fully separated. Hence, the contact force should vanish. This feature is well captured by G-W and the proposed MBMA model. However, it can be seen from Fig. 11 that the two Persson models still return contact forces of non-zero values (i.e., $F/A_0 \approx 3.5 \times 10^{-4} E^*$), when the two surfaces are separated from each other. This is because the asperities considered in Persson model have a non-zero probability of being arbitrary height (Yang and Persson, 2008a, 2008b; Papangelo et al. 2017). Such unrealistic contact force values (at $u = 3 h_{\text{rms}}$) may result in considerable engineering deviation. For example, if the two contact solids are made of steel, this non-zero contact force can reach as large as $F/A_0 = 70\text{MPa}$. Actually, it is read from the top-right figure in Fig. 11 that, when $u/h_{\text{rms}} > 2$, the computed value of $F/A_0$ by Persson models are substantially over-estimated. For instance, when $u = 2.5 h_{\text{rms}}$, the two Persson models suggest $F/A_0 \approx 1 \times 10^{-3} E^*$, which is about an order of magnitude higher than that predicted by the proposed MBMA model ($F/A_0 \approx 2.2 \times 10^{-4} E^*$). In fact, an accurate prediction of the actual contact force under relatively light indentation (i.e., $u$ is relatively large) is of great value in engineering applications. An example can be found is the case of leakage analysis of bolted flange joints without gaskets. The range of $u/h_{\rm rms} > 2$, (which corresponds to $F/A_0 < 2.9 \times 10^{-3} E^*$) plays a crucial role in its sealing performance (Ma et al., 2017). On the other end, if the full contact condition is met (i.e., $u = 0$), good matching results with the MD simulations are found in the two Persson models and the proposed MBMA model. In contrast, G-W model gives a prediction which is several orders magnitude higher than the MD results. Therefore, with the proposed MBMA model, the contact force during the entire compression process can be well captured as shown in Fig. 11. Moreover, the proposed MBMA model well connects the G-W model (which is believed valid for small indentation) and the Persson model (which is envisaged to hold true for large indentation).

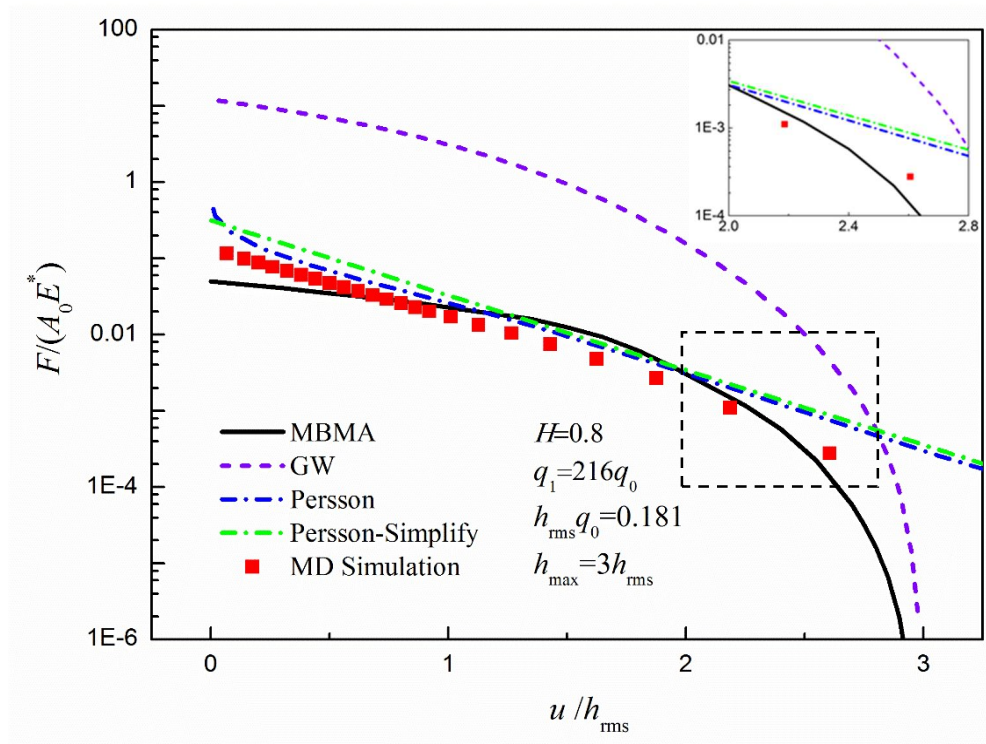

Fig. 11. Comparison with MD results: the actual contact force with respect to the average interfacial separation.

Another issue that is worth pointing out is that when comparison is made between G-W and Persson models, the following two quantities are generally treated as being equivalent: the distance between the rigid plane and the mid-plane of the rough surface $d$ and the average interfacial separation $u$ (Carbone and Bottiglione, 2011; Lorenz et al., 2010). However, such treatment may not be appropriate, as the two quantities are defined with different physical meanings. In Appendix D, the difference between $d$ and $u$ is discussed in detail, and an appropriate relation between $d$ and $u$ should be given by

$$u = (h_{\max} + d)/2, \tag{50}$$

where $h_{\max}$ is recalled to be the highest height of the asperities on the rough surface. It can be found that the relation $u = d = h_{\max}$ holds only at the initial contact state. On the other hand, when it becomes fully contacted $d = -h_{\max}$ in G-W model (which corresponds to the situation when the lowest asperity is compressed), correspondingly in Persson model, $u$ should become zero.

**4.1.2. Green-function-based and G-W discrete results**

The proposed MBMA model is also compared with the simulation results by Papangelo et al. (2017). In their works, both Green-function-based and G-W discrete approaches were adopted. The morphological parameters of the rough surface are taken as follows: Hurst exponent $H = 0.8$, the long distance roll-off wavenumber $q_0 = 2\pi/100 \times 10^6 \text{m}^{-1}$; the largest measured wavenumber $q_1 = 64 q_0$; the mean square root roughness $h_{\text{rms}} = 5 \times 10^{-6} \text{m}$ (Papangelo et al., 2017). Although not mentioned in the original work, the highest height of the asperities is set to be $h_{\max} \approx 3.0 h_{\text{rms}}$ by following conventions.

The simulation results are shown in Fig. 12. It is found that the MBMA model can accurately estimate the relationship between the contact force and the interfacial separation during the entire compression process. In addition, there may be slight deviation at the full contact status (when $u \to 0$), but the actual disparity is no more than 2 times, compared to the simulation results. This deviation may arise from the highest height of the rough surface is fixed to be $h_{\max} = 3.0 h_{\text{rms}}$, which limits the compressive strength of each asperity. Besides, it is also noted that the blue dotted line is the estimation given by the "G-W discrete" method (Ciavarella. et al., 2004).

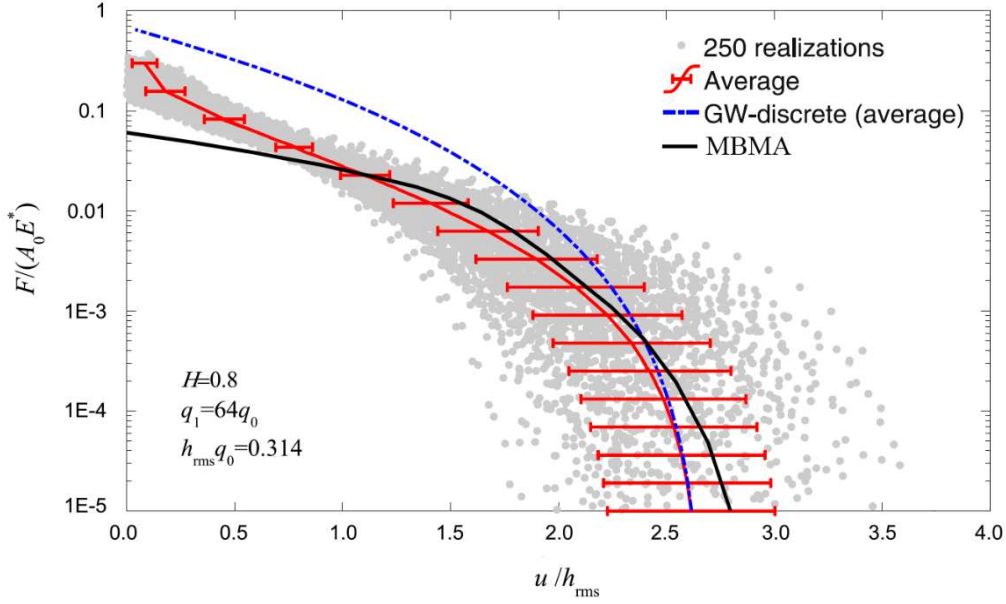

Fig. 12. Comparison with Green-function-based and G-W discrete results by Papangelo et al., (2017).

## 4.2. Comparison with the experimental results
### 4.2.1. Experimental results by Lorenz et al.

Recently, Lorenz et al. experimentally investigated the relationship between the contact force and the average interfacial separation (Lorenz, 2012; Lorenz et al., 2010). The experiment has been carried out by pressing a silicon rubber block against two different road surfaces: concrete road and asphalt road in displacement-control process. The morphological parameters for the concrete road surface are taken as follows: Hurst exponent $H = 0.7$, and the quantities $m_0 = 0.071 \text{mm}^2$, $m_2 = 0.27$, $m_4 = 62 \text{mm}^{-2}$. The highest height of the surface from the mid-plane is $h_{\max} = 1.1\text{mm}$, and thus the actual contact starts at $u = 4.1 h_{\text{rms}}$.

With these parameters, the contact status predicted by G-W model, Persson as well as Persson simplified model and the proposed MBMA model is respectively investigated. The corresponding results are also compared with experiment as shown in Fig. 13(a). According to the experimental data, the compression of the concrete road starts when the average interfacial separation equals $4.1h_{\mathrm{rms}}$, i.e., $u = 4.1h_{\mathrm{rms}}$. At this stage, the experimental value for the contact forces should drop to zero. Again, this phenomenon is well captured by the proposed MBMA approach and the G-W model. However, as the indentation strengthens, the result given by G-W model starts to deviate from the experimental data, while the MBMA model still matches the experimental observation well, along with the two Persson models. Therefore, the proposed MBMA model delivers a reasonably good prediction to the experiment throughout the entire compression process, when compared to the other two types of methods. Again, we notice the interesting feature displayed by the MBMA model: it well connects the G-W-type theory and Persson's model.

Similar conclusions can also be made when referring to the indentation of an asphalt road (Fig. 13(b)) where the morphological parameters are taken as follows: Hurst exponent $H = 0.85$, and the quantities $m_0 = 0.091\mathrm{mm}^2$, $m_2 = 2.67$, $m_4 = 21000\mathrm{mm}^{-2}$. The highest height of the surface from the mean plane is $h_{\max} = 1.37\mathrm{mm}$, and thus the actual contact starts at $u = 4.5h_{\mathrm{rms}}$.

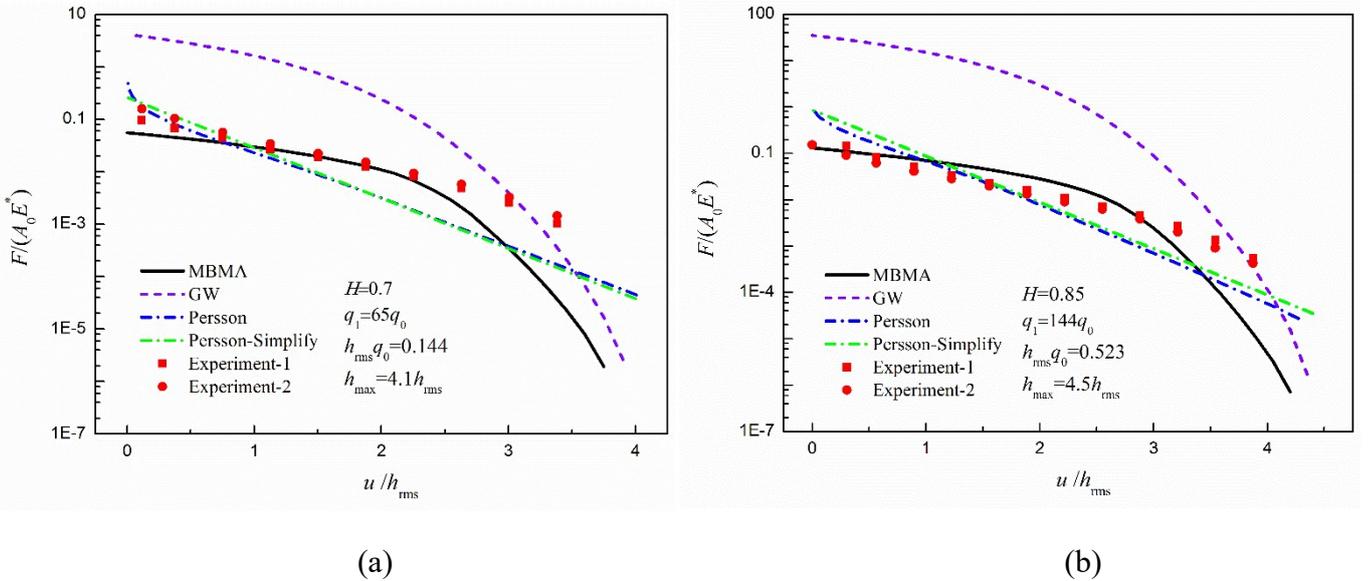

Fig. 13. The plot of the contact force with respect to the average interfacial separation for the contact between a rubber block and two different types of road samples: (a) concrete road surface, (b) asphalt road surface.

**4.2.2. Experimental results by Putignano et al.**

Putignano et al. (2013) conducted an experimental investigation on the real contact area respect to the contact force on the same concrete road sample as used by Lorenz et al. (Lorenz, 2012; Lorenz et al., 2010). It is known that the accurate prediction of the contact area, is crucial for many engineering applications, e.g., electrical and heat conductions, leakage in seals. Therefore, in Fig. 14(a), the predictions given by different models (i.e., the proposed MBMA model, the G-W and Persson models), are compared with the experimental results. It is observed that the predictions obtained by the Persson and the proposed MBMA models match the experiment results well, while the result given by G-W model underestimates the value of the contact area with a given contact force. In addition, the asphalt road sample in (Lorenz, 2012; Lorenz et al., 2010) is also employed to theoretically calculate the contact area with respect to the contact force, and the results are shown in Fig. 14(b). It is found that similar trends can be seen as those obtained with the concrete road sample.

Moreover, it has been pointed out that under small indentation conditions, a linear relation (Campaná and Mueser, 2007; Carbone and Bottiglione, 2008; Ciavarella et al., 2006b; Hyun et al., 2004; Persson, 2006; Putignano et al., 2012; Putignano et al., 2013; Yastrebov et al., 2017):

$$A = \frac{\kappa}{\sqrt{2m_2}E^*} \cdot F \tag{51}$$

should hold between the contact area $A$ and the contact force $F$, where the scaling factor $\kappa$ is suggested to be roughly 2 by many numerical predictions (finite element methods and molecular dynamic simulations) (Campaná and Mueser, 2007; Hyun et al., 2004; Putignano et al., 2012; Putignano et al., 2013). Meanwhile, it is found that the value of $\kappa$ is $\sqrt{2\pi}$ for G-W approach, and is $\sqrt{8/\pi}$ for Persson theory (Akarapu et al., 2011; Campaná and Mueser, 2007; Carbone and Bottiglione, 2008; Hyun et al., 2004; Persson, 2006; Putignano et al., 2012; Yastrebov et al., 2017). The proposed MBMA model is also employed to calculate $\kappa$ for the two cases shown in Fig. 14(a) and Fig. 14(b). For the concrete road sample, the obtained value of $\kappa$ is 1.98, and for asphalt road sample, 1.75, both of which are closer to 2 compared to that calculated by using G-W and Persson models.

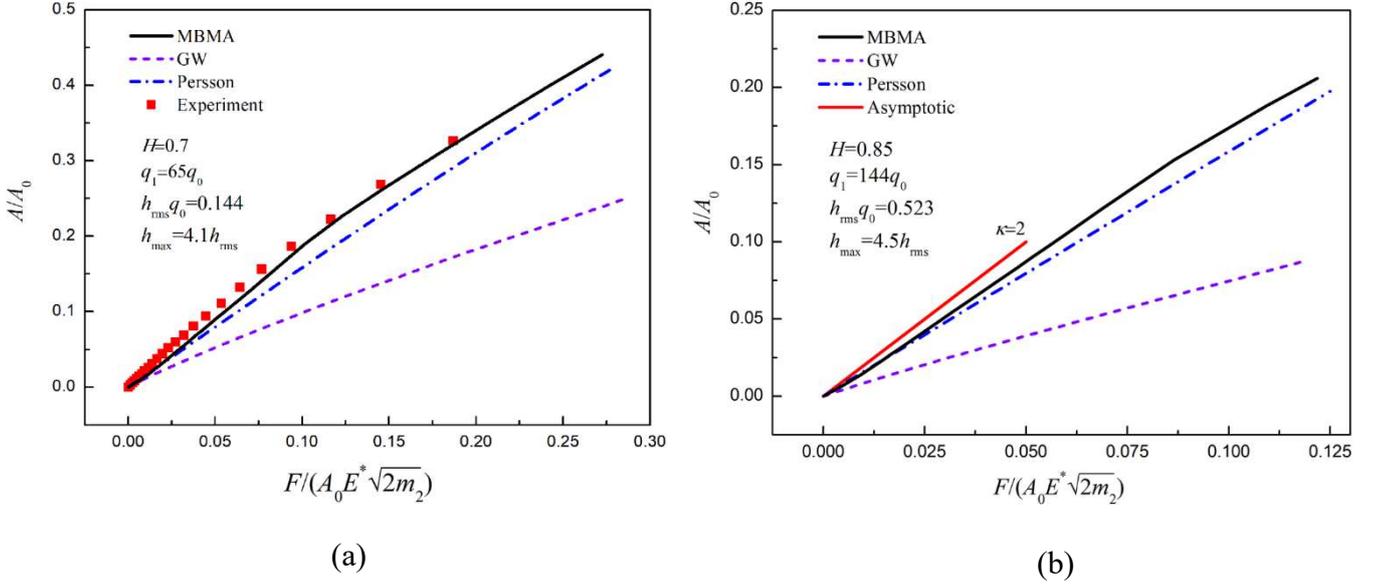

Fig. 14. The plot of the contact area with respect to the contact force for the contact between a rubber block and two different types of road samples: (a) concrete road surface, (b) asphalt road surface.

### 4.3. Comparison with "Contact challenge" organized by Mueser et al.

Recently, Mueser et al. (2017) proposed an "contact-mechanics modeling challenge", and many famous solution strategies (ranging from traditional asperity-based models via Persson theory and brute-force computational approaches, to real-laboratory experiments and all-atom molecular dynamics simulations) are invited to make predictions on a mathematically well-defined problem. The morphological parameters of the rough surface are taken as follows: Hurst exponent $H = 0.8$, the long distance roll-off wavenumber $q_0 = 2\pi/20 \times 10^6 \text{m}^{-1}$; the largest measured wavenumber $q_1 = 200 q_0$; the mean square root roughness $h_{\text{rms}} = 0.762 \times 10^{-6}\text{m}$; the highest height of the asperities $h_{\text{max}} \approx 3.7 h_{\text{rms}}$ (Mueser et al., 2017).

The proposed MBMA model is also employed without considering adhesion. As shown in Fig. 15(a), the relationship between the contact force and the average interfacial separation calculated by the proposed MBMA model is in a reasonable range compared with other solution strategies. Moreover, there may be slight deviation at the full contact status (when $u \to 0$), but the actual disparity is no more than 4 times, compared to other cluster of solution strategies (e.g., Persson approach, FFT-BVM and GFMD). In Fig. 15(b), the proposed MBMA model is also used to plot the real contact area as a function of the contact force. Throughout the entire compression process, the results estimated by the MBMA model are quite accurate, especially at the full contact status (i.e., $u = 0$). It is worth noting that under small contact force, the calculated contact area by the MBMA model is a little bit lower than that associated with most of other solution strategies. This deviation is understood due to the fact that the proposed MBMA model is derived in the absence of adhesion. Overall speaking, the proposed MBMA model performs well in this "contact challenge" test.

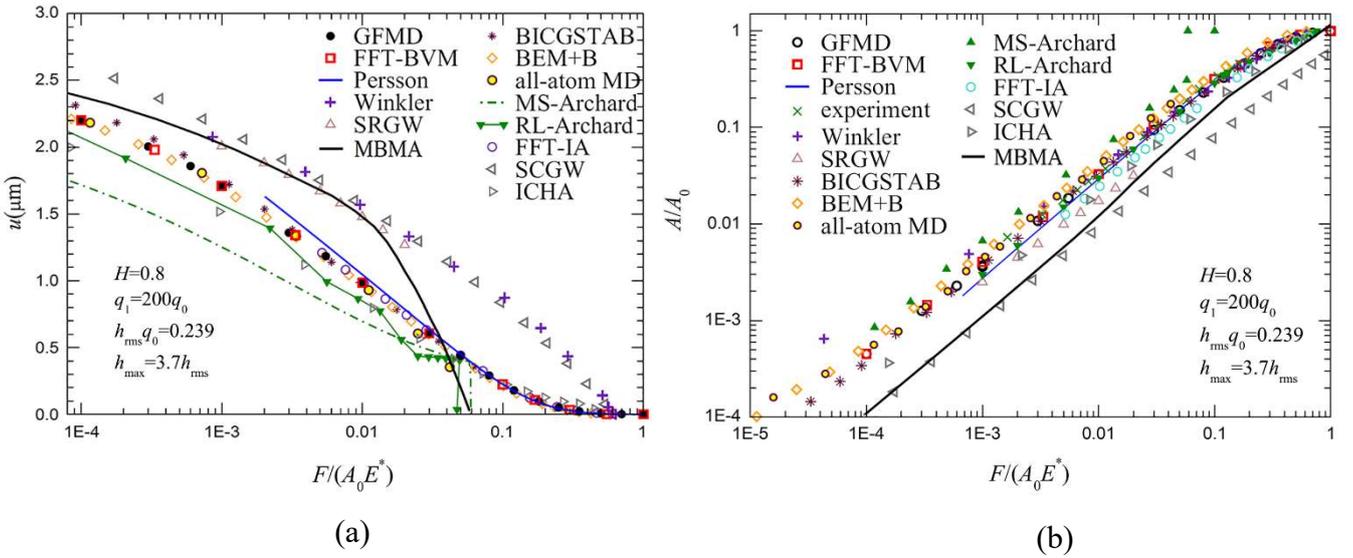

Fig. 15. Comparison with the data from the "contact challenge" test coordinated by Mueser et al. (2017). The plot of the contact force with respect to: (a) average interfacial separation, (b) contact area.

### 4.4. Sensitivity to the measuring resolution of rough surfaces

In theory, a real surface is a self-affine fractal spanning over a wide range of length scales with $H > 0.7$ (Persson, 2014), and the shortest wavelength can be as small as atomic distances (Persson, 2006; Yang and Persson, 2008a). However, the present identification technique still finds it difficult to resolve all length scales associated with rough surfaces. From that viewpoint, we will check how the existing rough-surface models depend on the truncation wavenumber $q_1$ (when constructing the $C(q)$). The morphological parameters of the rough surface are chosen as follow: Hurst exponent $H = 0.7$, and $q_0 h_{\text{rms}} = 0.1$. The highest height of the surface from the mean plane is fixed to be $h_{\max} = 3 h_{\text{rms}}$, and the largest measured wavenumber $q_1$ are 125, 250, 500, 1000 times more than the smallest measured wavenumber $q_0$, respectively. The contact force with respect to the average interfacial separation calculated by G-W model, Persson simplified model and the proposed MBMA method are plotted in Fig. 16.

It can be observed from Fig. 16 that for small indentations (i.e., when $u$ is large), only the G-W and the MBMA models are sensitive to $q_1$. Such sensitivity at small indentation is also reported in Yang and Persson's (2008a; 2008b) MD simulation. In fact, the sensitivity to the value of $q_1$ can be rationalized through the derivation of the MBMA model, where asperities of large curvature radii barely deform under small indentations. Thus, it is those small-scale asperities that determine the actual contact status of a rough surface when the indentation is small. For large indentations, the spectrum of asperities that play a key role in withstanding indentations move leftward in $q$-axis (see Fig. 2 for reference). Hence the actual contact force should become less reliable on small-scale wavenumbers, or $q_1$. This feature is also seen in the outcomes obtained by both the Persson model and the proposed MBMA method. In contrast, the result by G-W model is still sensitive to $q_1$ under high indentation (for small $u$). This is because G-W model is based on an average over the whole asperity spectrum. Thus an increase in $q_1$ results in a decline in the mean curvature radius of asperities. Therefore, it is likely to conclude that the contact force is highly sensitive to $q_1$ at small indentations, and becomes less sensitive at large indentation. Again, this finding can also be verified by Yang and Persson's (2008a; 2008b) MD simulation and the numerical results from Putignano et al. (2012).

In Fig. 17, the contact force-area relationships with various $q_1$ are generated based on the three aforementioned methods. It appears that compared with the predictions obtained by G-W model (which are much lower than the well-known scaling factor $\kappa = 2$), the proposed MBMA model is more accurate during the entire compression process with arbitrary value of $q_1$.

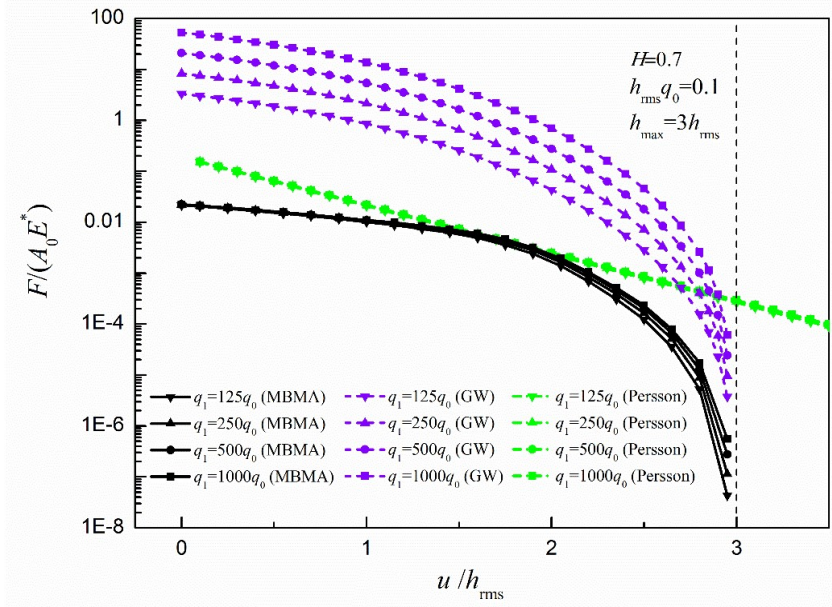

Fig. 16. The contact force with respect to the average interfacial separation with different detected largest wavenumber $q_1$ calculated by G-W model, Persson's simplified model and the proposed MBMA model.

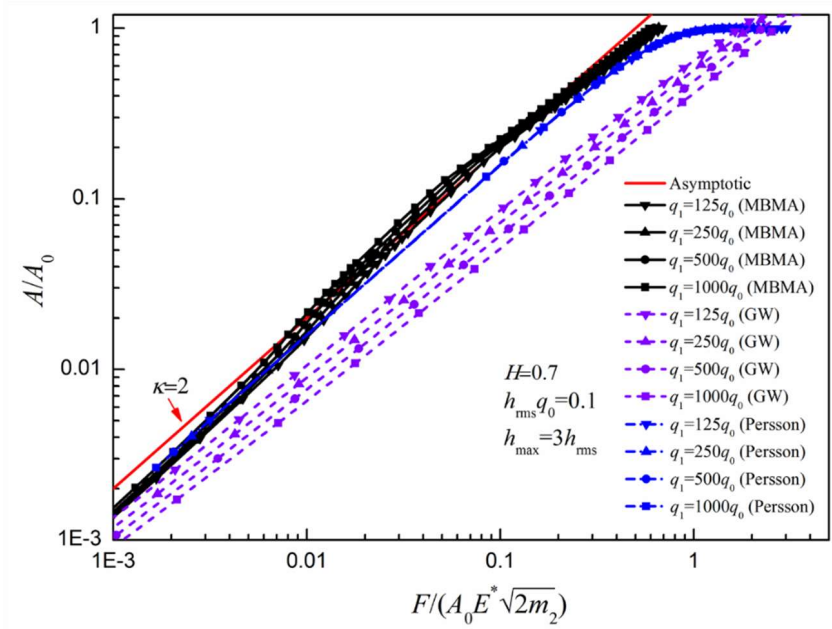

Fig. 17. The real contact area with respect to the contact force with different detected largest wavenumber $q_1$ calculated by G-W model, Persson theory and the proposed MBMA model.

## 5. Conclusion

By integrating the magnification quantitative method of rough surfaces introduced by Persson into the asperity-representation presented by Greenwood and Williamson, a magnification-based multi-asperity model (which gives a deeper understanding of the contact behavior) for analyzing the contact status between rough surfaces is established in this paper. The key idea underpinning the MBMA model is to transform the originally complex contact problem to a family of sub-contact problems on "contact islands", each of which can be illustrated as a sub-tree structure as shown in Fig. 6. Then for a given indentation distance, only asperities (associated with the $i$-th contact island) whose radii fall into an effective range (i.e., $R_d^i$ and $R_u^i$) play a major role in withstanding the indentation, and this effective range will evolve accordingly as the indentation intensifies. The derived MBMA model is expected to deliver a relatively accurate prediction to the actual contact status throughout the entire compression process, and this has been verified by comprising with theoretical, numerical and experimental data. Besides, it is shown that the two families of widely used theories, i.e., G-W model, and Persson models are well connected by the proposed MBMA model. Moreover, the scale effect induced by the resolutions of measurement, which has been well observed in MD simulations (Yang and Persson, 2008a; Yang and Persson, 2008b), is effectively captured by the proposed MBMA model. In addition, it is worth noting that, the classical elastic iterative algorithm proposed by Greenwood and Tripp (1967) is imposed in the MBMA model so as to ensure the accuracy of computing results for the individual contact island. However, the implantation of such iterative algorithms inevitably brings down the computational efficiency. In order to save computational time, one may refer to the work by Bahrami et al. (2005) who have summarized an empirical expression formulating the plastic deformation of spherical rough contacts. In a similar way, empirical formulation of the elastic behavior under the set-up given by Fig. 7(b) may be derived, and this indicates one possible research direction for future work. Besides, the idea underlying the derivation of the proposed MBMA model can also be generalized to studying the contact problems, where plasticity, leakage, and adhesion are considered. These research topics can be pursued in future works as well.

## Acknowledgements

The financial supports from the National Key Research and Development Plan (2016YFB0201600, 2016YFB0201601) from the Ministry of Science and Technology of the People's Republic of China, National Natural Science Foundation of China (11732004, 11772076), the Fundamental Research Funds for the Central Universities (DUT16RC(3)091), Program for Changjiang Scholars, Innovative Research Team in University (PCSIRT) are gratefully acknowledged.

**Appendix A:** Notation

In this appendix, notations that have been frequently used throughout the article are listed here.

$A, A_0$ - actual and nominal contact area

$A^i$ - the actual contact area on the $i$-th contact island

$C(q)$ - the power spectral density function of a rough surface

$d$ - distance between a rigid flat surface and the mid-plane of the original rough surface (see Fig. 3 and Fig. 5)

$D_{\text{sum}}$ - surface asperity density

$D_{\text{sum}}^i$ - the density of asperities whose characteristic wavenumbers falling in the range of $[q_0 \zeta_{\text{d}}^i, q_0 \zeta_{\text{u}}^i]$

$D_{\text{sum}}^d$ - surface asperity density observed at magnification $\zeta_{\text{d}}^i$

$D_{\text{sum}-\text{t}}$ - surface asperity density observed at the largest magnification $\zeta_1$

$E^*$ - equivalent Young's modulus between two contact surfaces

$E_1, E_2$ - Young's modulus of the two contact surfaces

$F$ - total contact force on the rough surface

$F^i$ - contact load on the $i$-th contact island

$H$ - Hurst exponent

$\underline{h}$ - mean square root roughness whose characteristic wavenumbers fall in the range of $[q_0, q_0\zeta]$ (see Eq. (24))

$\overline{h}$ - mean square root roughness whose characteristic wavenumbers fall in the range of $[q_0\zeta, q_1]$ (see Eq. (25))

$\overline{h}_{\text{d}}$ - mean square root roughness whose characteristic wavenumbers fall in the range of $[q_0\zeta_{\text{d}}^i, q_1]$

$\overline{h}_{\text{u}}$ - mean square root roughness whose characteristic wavenumbers fall in the range of $[q_0\zeta_{\text{u}}^i, q_1]$

$h_{\text{rms}}$ - mean square root roughness of the rough surface

$h_{\text{rms}}^i$ - mean square root roughness whose characteristic wavenumbers fall in the range

of $[q_0\zeta_d^i, q_0\zeta_u^i]$

$h_{\max}$- maximum height of the asperities on the rough surface observed at magnification $\zeta_1$ (see Fig. 5)

$k$ - stiffness of a rough surface consisting of asperities whose characteristic wavenumbers fall in the range of $[q_0\zeta_d^i, q_1]$

$K$- stiffness of the spherical asperity of radius $R_d^i$

$m_n$- the $n$-th Nayak's moments (see Eqs. (1) and (7))

$m_n(\zeta)$- the $n$-th Nayak's moments observed at magnification $\zeta$

$p, p_0$- actual and nominal contact stress

$q$- wavenumber

$q_0$- long distance roll-off wavenumber

$q_1$- the largest wavenumber that is detectable

$R$- average asperity summit curvature radius

$R_u^i$- the shortest curvature radius of all asperities that play an effective role in determining the contact status of the $i$-th contact island (see Fig. 5 and Fig. 8)

$R_d^i$- the largest curvature radius of all asperities that play an effective role in determining the contact status of the $i$-th contact island (see Fig. 5 and Fig. 8)

$R_{\text{top}}$- average summit curvature radius observed at magnification $\zeta_1$

$u$- average interfacial separation between two contact rough surfaces (see Fig. 18)

$z(x)$- surface height measured at $x$

$z_{\max}^i$- distance between the highest point of the $i$-th contact island (observed at the largest magnification $\zeta_1$) and its mean plane (see Fig. 8)

$\alpha$- Nayak parameter (see Eq. (3))

$\lambda$- wavelength

$\nu_1, \nu_2$- Poisson ratios of the two contact surfaces

$\sigma$- standard deviation of the summit heights

$\sigma_d$- standard deviation of the summit heights observed at magnification $\zeta_d^i$

$\sigma^i$- standard deviation of the summit heights of the roughness falling in the range of $[q_0\zeta_d^i, q_0\zeta_u^i]$

$\delta^i$ - indentation distance on the $i$-th contact island (see Fig. 8)

$\delta_\text{u}$ - indentation distance of the highest sphere asperity observed at magnification $\zeta_\text{u}^i$ into the $i$-th contact island (see Fig. 8)

$\eta$ - the ratio of the real contact area (when consider the ignoring roughness in the range of $[q_0\zeta_\text{u}^i, q_1]$) to its Hertz contact area on the spherical asperities with curvature radius $R_\text{u}^i$

$\zeta$ - current magnification (see Eq. (13))

$\zeta_1$ - measured largest magnification

**Appendix B**: Estimation of $R_d^i$ and $R_u^i$

In this appendix, the two bounds for the curvature radii of asperities in the $i$-th contact island (i.e., $R_u^i$ and $R_d^i$ defined in Section 3.2) are derived in detail.

**B.1.** Estimation of $R_d^i$

Referring to the results given by Phort and Popov (2013), where the contact status between an elastic rough sphere and a rigid plane is investigated by using three-dimensional boundary element methods. They pointed out that when the stiffness of an elastic sphere (denoted by $K$) is greater than that induced by its surface roughness (denoted by $k$), i.e,

$$k \leq K, \tag{B.1}$$

it is the roughness of the sphere that dominates the deformation. In order to employ this principle, we calculate $k$ as the stiffness of the upper elastic surface in Fig. 7(b) by (Akarapu et al., 2011)

$$k = 2F^i/\overline{h}_d, \tag{B.2}$$

where $F^i$ is the contact force imposed on the $i$-th contact island; $\overline{h}_d$ is the mean square root roughness whose characteristic wavenumbers fall into the range of $[q_0\zeta_d^i, q_1]$. The expression for $\overline{h}_d$ can be calculated by replacing $\zeta$ in Eq. (25) by $\zeta_d^i$:

$$\overline{h}_d = h_{\text{rms}}\sqrt{\frac{(\zeta_d^i)^{-2H} - \zeta_1^{-2H}}{1 - \zeta_1^{-2H}}}. \tag{B.3}$$

On the other hand, the stiffness of the sphere $K$ can be calculated by employing the Hertz formulation (Johnson, 1987):

$$K = \left(6R_d^i F^i E^{*2}\right)^{1/3}. \tag{B.4}$$

Substituting Eqs. (B.2) and (B.4) into (B.1), we obtain

$$2F^i \leq \overline{h}_d\left(6R_d^i F^i E^{*2}\right)^{1/3}. \tag{B.5}$$

One way to calculate the contact force $F^i$ is by employing

$$F^i = \frac{4}{3}E^*\left(R_d^i\right)^{\frac{1}{2}}(\beta\delta^i)^{\frac{3}{2}}, \tag{B.6}$$

where $\beta \in (0,1)$ is the ratio of deformation assigned to the elastic spheres (shown in Fig. 7(b)) from the total indentation distance $\delta^i = z^i_{\max} - d$. Substituting Eq. (B.6) into (B.5), we find that

$$\frac{4}{3}\beta(z^i_{\max} - d) \leq \overline{h}_\mathrm{d}. \tag{B.7}$$

Incorporating Eq. (B.3) into (B.7) gives

$$\zeta^i_\mathrm{d} \leq \left\{\left[\frac{4}{3}\beta\left(\frac{z^i_{\max} - d}{h_{\mathrm{rms}}}\right)\right]^2 + \zeta_1^{-2H}\right\}^{-\frac{1}{2H}}. \tag{B.8}$$

In fact, under the regime where Eq. (B.1) holds, the elastic rough surface in Fig. 7(b) should deform more than the elastic surface consisting of spheres of radius $R^i_d$, i.e., $\beta < 1/2$. Here we set $\beta$ to be $1/3$ (the influence of $\beta$ falling in the value range $[1/3, 1/2]$ on the calculated results of contact status is very small), which can ensure the calculated value of $\zeta^i_\mathrm{d}$ satisfying Eq. (B.1) all the time. Thus Eq. (B.8) becomes

$$\zeta^i_\mathrm{d} \leq \underline{\zeta}^i = \left\{\left[\frac{4}{9}\left(\frac{z^i_{\max} - d}{h_{\mathrm{rms}}}\right)\right]^2 + \zeta_1^{-2H}\right\}^{-\frac{1}{2H}}. \tag{B.9}$$

In theory, the lower bound for $\zeta^i_\mathrm{d}$ should be $\zeta = 1$, when no asperities are observable. However, letting $\zeta^i_\mathrm{d} = 1$ may result in singular behavior in the computation of the proposed model. Hence, based on the simulation results to be presented later, we suggest that the lower bound for $\zeta^i_\mathrm{d}$ is 1.5, i.e.,

$$\zeta^i_\mathrm{d} = \max\left(1.5, \underline{\zeta}^i\right). \tag{B.10}$$

The value of $R^i_\mathrm{d}$ is then calculated by

$$R^i_\mathrm{d} = \frac{3\sqrt{\pi}}{8\sqrt{m_4(\zeta^i_\mathrm{d})}}, \tag{B.11}$$

where $m_4(\zeta^i_\mathrm{d})$ is evaluated based on Eq. (23).

**B.2.** Estimation of $R^i_\mathrm{u}$

According to the conclusion by Greenwood and Tripp (1967) and Johnson (1987), it is known that the non-dimensional parameter $\varphi$ calculated based on Eq. (28) is crucial to determine the values of $R^i_\mathrm{u}$ and $\zeta^i_\mathrm{u}$. Referring to the situation considered here, $\sigma_{\mathrm{u}1}$ can be approximated by the mean square root roughness of the upper elastic rough surface shown in Fig. 7(b), which is quantified by $\overline{h}_u$ here, and $\overline{h}_u$ is calculated by replacing $\zeta$ in Eq. (25) by $\zeta_u^i$:

$$\sigma_{u1} \approx \overline{h}_u = h_{rms}\sqrt{\frac{(\zeta_u^i)^{-2H} - \zeta_1^{-2H}}{1 - \zeta_1^{-2H}}}. \tag{B.12}$$

To evaluate the denominator $\delta_u$ in Eq. (28), for simplicity, we mainly consider the deformation of the highest spherical asperity as shown in Fig. 7(b). Since the highest point is as high as $z_{max}^i - \overline{h}_u$ before compressed, we thus have

$$\delta_u = z_{max}^i - d - \overline{h}_u, \tag{B.13}$$

where $z_{max}^i$ is the highest height of the original rough surface in the $i$-th contact island observed at the magnification $\zeta_1$. A schematic illustration of the derivation of Eq. (B.13) is seen in Fig. 8.

Combining the Eqs. (28) to (B.13) and letting $\varphi \leq 0.05$, gives

$$\overline{h}_u \leq \frac{z_{max}^i - d}{21}. \tag{B.14}$$

By employing the expression for $\overline{h}_u$ in Eq. (B.12), we solve the inequality (B.14) to obtain

$$\overline{\zeta}^i \leq \zeta_u^i \leq \zeta_1, \tag{B.15a}$$

where

$$\overline{\zeta}^i = \left[\left(\frac{z_{max}^i - d}{21}\right)^2 \cdot \frac{1 - \zeta_1^{-2H}}{h_{rms}^2} + \zeta_1^{-2H}\right]^{-\frac{1}{2H}}. \tag{B.15b}$$

It is recalled that the lowest value of $\zeta_d^i$ has been set to be 1.5. To prevent the value of $\zeta_u^i$ from approaching 1.5, we require that $\zeta_u^i$ should not fall below 3. Hence, we have

$$\zeta_u^i = \max\left(3, \overline{\zeta}^i\right). \tag{B.16}$$

Then, $R_u^i$ can be calculated as

$$R_u^i = \frac{3\sqrt{\pi}}{8\sqrt{m_4(\zeta_u^i)}}, \tag{B.17}$$

where $m_4(\zeta_u^i)$ is evaluated based on Eq. (23).

**Appendix C**: Calculation of $\eta$ in Eq. (39)

Here, the focus is laid on one of the spherical asperities with the curvature radius $R_u^i$ as depicted by the red box in Fig. 9. In the derived model, the roughness in the range $[q_0 \zeta_u^i, q_1]$ of the original rough surface has been ignored as they have been flattened, and it has been shown by numerical results that the calculation of the compression force is barely affected upon this simplification. However, this treatment may significantly influence the calculation of the real contact area. As a remedy, here an influence coefficient $\eta$ which defined as the ratio of the real contact area $A_u$ (when consider the above ignoring roughness) to its Hertz contact area $A_\text{Hertz}$ on the spherical asperities with the curvature radius $R_u^i$, i.e.,

$$\eta = \frac{A_u}{A_\text{Hertz}} \tag{C.1}$$

is introduced to help capture the actual contact area. In Eq. (C.1), $A_\text{Hertz}$ is calculated as

$$A_\text{Hertz} = \pi L_\text{H}^2, \tag{C.2}$$

and $L_\text{H}$ is the size of the Hertz contact zone, which should satisfy

$$L_\text{H} = \sqrt{\delta_u R_u^i} \tag{C.3}$$

according to the Hertz (1881) contact theory. The symbol $\delta_u$ appeared in Eq. (C.3) is the indentation of the highest sphere asperity with radius $R_u^i$ shown in Fig. 8. With reference to Section 3.2, $\overline{h}_u/\delta_u = 0.05$. Inserting this into Eq. (C.3) gives

$$L_\text{H} = \sqrt{20 \overline{h}_u R_u^i}, \tag{C.4}$$

where $\overline{h}_u$ (see Eq. (B.12)) is recalled to be the mean square root roughness of the rough surface whose characteristic wavenumbers fall into the range of $[q_0 \zeta_\text{up}^i, q_1]$.

On the other hand, $A_u$ is calculated by summarizing the small contact area $a'$ of an individual small asperity of the ignoring roughness whose characteristic wavenumbers fall into the range of $[q_0 \zeta_\text{up}^i, q_1]$:

$$A_{\mathrm{u}} = \int_0^{L_{\mathrm{H}}} a' D_{\mathrm{sum-t}} 2\pi r \mathrm{d}r, \tag{C.5}$$

and

$$a' = \pi \left(\frac{3fR_{\mathrm{top}}}{4E^*}\right)^{\frac{2}{3}}. \tag{C.6}$$

In Eqs. (C.5) and (C.6), $D_{\mathrm{sum-t}}$ and $R_{\mathrm{top}}$ are the surface asperity density and the average summit curvature radius relating to the magnification $\zeta_1$, which can be obtained from Eqs. (2), (6) and (7). In Eq. (C.6), $f$ is the load that an individual asperity of the ignoring roughness undertakes, and it can be calculated as

$$f = p_r/D_{\mathrm{sum-t}}, \tag{C.7}$$

where

$$p_r = \bar{p}_0 \sqrt{1 - \left(\frac{r_1}{L_{\mathrm{H}}}\right)^2} \tag{C.8}$$

is the contact pressure on the sphere asperity with radius $R_{\mathrm{u}}^i$; $\bar{p}_0$ denote the maximum stress, which can be calculated as

$$\bar{p}_0 = \frac{2E^* L_{\mathrm{H}}}{\pi R_{\mathrm{u}}^i} \tag{C.9}$$

based on the Hertz (1881) contact theory. Therefore, combining Eqs. (C.1) to (C.9), the value of $\eta$ is

$$\eta = \frac{3}{4}\pi^{1/3} D_{\mathrm{sum-t}}^{1/3} \left(\frac{3R_{\mathrm{top}}}{2}\right)^{\frac{2}{3}} \left(\frac{20\bar{h}_{\mathrm{u}}}{R_{\mathrm{u}}^i}\right)^{\frac{1}{3}}, \tag{C.10}$$

which appeared in Eq. (39).

**Appendix D**: The transformation between $d$ and $u$

In G-W model, the compression is measured in terms of $d$, which is the distance between the lower rigid surface and the original mid-plane of the rough surface, while in Persson model, it is measured in terms of $u$, which is the average interfacial separation as shown in Fig. 18. Referring to Fig. 18, where a rough surface is compressed by a flat rigid surface under a contact force $F$, it is observed that

$$u = d + w, \qquad (D.1)$$

where $w$ is the relative displacement of the mid-plane of the rough surface. As suggested by (Lorenz, 2012; Lorenz et al., 2010; Wang et al., 2014), the average interfacial separation $u$ is calculated by

$$u = h_{\max} - w. \qquad (D.2)$$

Combining Eqs. (D.1) and (D.2) gives

$$u = (h_{\max} + d)/2, \qquad (D.3)$$

which is Eq. (50).

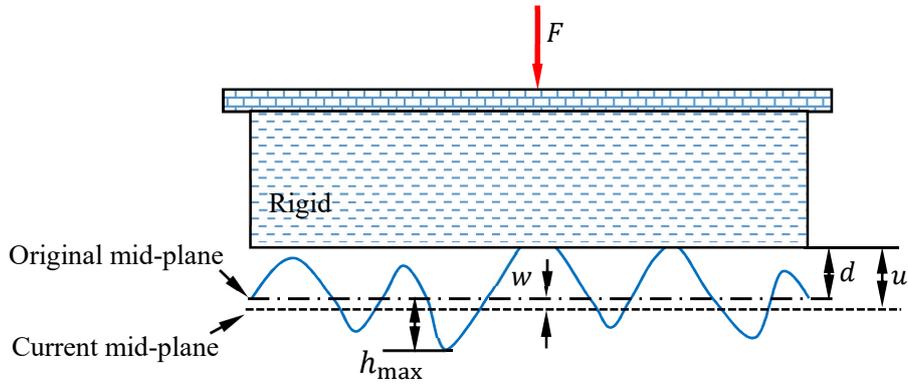

Fig. 18. The contact between a rigid surface and an elastic solid with rough surface.